\documentclass[journal]{IEEEtran}

\usepackage[final]{changes} 

\usepackage{amsmath,amsfonts}
\usepackage{algorithmic}
\usepackage{array} 
\usepackage{textcomp}
\usepackage{stfloats}
\usepackage{url}
\usepackage{verbatim}
\usepackage{graphicx}
\def\BibTeX{{\rm B\kern-.05em{\sc i\kern-.025em b}\kern-.08em
    T\kern-.1667em\lower.7ex\hbox{E}\kern-.125emX}}
\usepackage{balance}

\usepackage{cite}
\usepackage{booktabs} 
\usepackage{makecell}
\usepackage{tabularx}
\usepackage{threeparttable}
\usepackage{caption}
\usepackage{array}

\newcolumntype{C}{>{\centering\arraybackslash}X}
\newcolumntype{L}{>{\raggedleft\arraybackslash}X}
\newcolumntype{R}{>{\raggedright\arraybackslash}X}

\newcommand{\secref}[1]{Section~\ref{#1}}
\newcommand{\fgref}[1]{Fig.~\ref{#1}}
\newcommand{\tabref}[1]{Table~\ref{#1}}

\usepackage{subfigure}
\usepackage{subcaption}
\usepackage[ruled, linesnumbered, vlined]{algorithm2e}  
\usepackage[colorlinks,linkcolor=red,anchorcolor=blue,citecolor=green]{hyperref}

\begin{document}
\title{
STISA: A 0.16 GOPS/W/PE Single-Shot Inference FPGA-based SNN Accelerator with Algorithm and Hardware Co-Design
}
\author{Kainan Wang, Chengyi Yang, Lei Liu, Chengting Yu, \\Yee Sin Ang, Bo Wang,~\IEEEmembership{Senior Member,~IEEE} and Aili Wang,~\IEEEmembership{Member,~IEEE} 
\thanks{
This work was supported in part by National Natural Science Foundation of China under grant 62304203, in part by International Campus of ZJU international research collaboration seed project, and in part by Natural Science Foundation of Zhejiang Province, China under Grant ZCLZ24F0102. 
\textit{(Corresponding author: Aili Wang.)} 
}
\thanks{Kainan Wang, Chengyi Yang, Lei Liu, Chengting Yu, and Aili Wang are with the Zhejiang University-University of Illinois Urbana-Champaign Institute, Zhejiang University, Haining, Zhejiang China. (e-mail: \{kainan.22, chengyi.22, chengting.21, lei.23, ailiwang\}@intl.zju.edu.cn).}

\thanks{Yee Sin Ang and Bo Wang are with the Information Systems Technology and Design,
Singapore University of Technology and Design, Singapore (e-mail: \{yeesin\_ang, bo\_wang\}@sutd.edu.sg).}
}

\maketitle

\begin{abstract} 
Spiking neural networks (SNNs) offer great potential for low-power intelligent computing owing to their event-driven nature, yet their deployment is limited by high inference latency and inefficient hardware utilization caused by temporal dependencies. Existing accelerators struggle to balance timesteps, accuracy, and energy efficiency, while hardware-oriented timestep compression remains underexplored. To address these challenges, this work presents STISA, a unified algorithm-hardware co-design framework featuring: (i) a Temporal Splitter Compression (TSC) technique with a Temporal Split Structure (TSS) to reduce temporal redundancy while preserving network dynamics; (ii) a hardware-efficient spatiotemporal parallelism scheme, which co-optimizes dataflow and architecture through a Temporal-Prioritized Output-Stationary (TPOS) mapping and a flexible Block-Mux Streaming (BMS) architecture; and (iii) a joint TSC-TPOS-BMS co-optimization framework featuring a bandwidth-aware resource allocation strategy to achieve balanced accuracy, latency, and energy efficiency across diverse network architectures. 
On the algorithmic side, TSC reduces synaptic operations by 24.44\%-51.71\% while achieving competitive accuracies of 96.38\%, 81.10\%, 68.51\%, and 82.40\% on CIFAR-10, CIFAR-100, ImageNet, and DVS-CIFAR10, respectively. On the hardware side, FPGA prototypes deliver up to 7.99× speedup, 30.4\% power reduction, and 30.1\% LUT reduction. Compared with state-of-the-art SNN accelerators, STISA achieves 0.16 GOPS/W/PE, demonstrating its scalability and superior energy efficiency for real-time SNN inference under constrained hardware resources.
\end{abstract}

\section{Introduction}   \label{introduction}
\IEEEPARstart{R}{ecently}, 
Spiking Neural Networks (SNNs) have emerged as a promising paradigm for energy-efficient intelligent computing due to their event-driven and bio-inspired temporal dynamics. However, the inherent temporal dependency of SNNs poses significant challenges to hardware realization: accurate inference often requires multiple timesteps, resulting in high latency, large memory footprints, and suboptimal hardware utilization. Balancing accuracy, latency, and energy efficiency therefore remains a central bottleneck in neuromorphic acceleration.

From the hardware perspective, existing SNN accelerators exhibit diverse design philosophies in handling temporal computation. Early works such as SpinalFlow \cite{Spinalflow} employed a temporal-sequential scheme, processing all timesteps of one layer before proceeding to the next, which preserves neuron-state locality but incurs strictly serial temporal execution within each layer. Parallel Time Batching \cite{PTB} partially alleviated this limitation by introducing partially temporal-parallel windows, where multiple systolic arrays process different segments of the inference sequence concurrently. However, timesteps within each window still execute sequentially, limiting the achievable latency reduction.

More recent works, including Yin et al. \cite{Loas}, Fang et al. \cite{fang2024jssc} and Zhang et al. \cite{zhang2025power}, further advanced fully temporal-parallel dataflows, where each timestep maintains an independent accumulator and membrane register. This design effectively eliminates temporal dependency and achieves significant latency reduction, yet incurs linearly increasing area and power overhead with the number of timesteps. Moreover, membrane-potential storage remains an energy bottleneck for longer time windows or multi-layer networks. Alternative approaches, such as the array-rearrangement strategy in \cite{zhang2025power} and the layer-wise streaming pipeline of DeepFire2 \cite{Deepfire2}, improve utilization or throughput but suffer from scalability issues, either due to rapid growth in hardware resources or limited temporal representation capability. In summary, existing architectures reveal a fundamental trade-off: increasing temporal parallelism reduces latency but exacerbates area, power, and storage costs. Achieving scalable spatiotemporal parallelism under limited hardware resources remains an open challenge.

From the algorithmic perspective, recent advances focus on temporal compression and adaptive timestep execution to reduce redundant computation. Methods such as DIET-SNN \cite{Diet} and Temporal Pruning \cite{chowdhury} achieve near-ANN accuracy with fewer timesteps but mainly target static vision tasks. Dynamic-timestep schemes, such as SEENN \cite{Seenn} and Input-aware Dynamic-Timestep SNN \cite{Iadyt}, further improve energy efficiency by adaptively adjusting the number of timesteps for each input. While they offer flexible energy-latency trade-offs, their irregular execution patterns resulting from sample-dependent timestep counts induce dynamically varying resource demands, complicating deterministic scheduling and parallel mapping in hardware.

Recent algorithm-hardware co-designs, such as STELLAR \cite{Stellar} and LCT-based redundancy reduction \cite{Xie} attempt to bridge this gap by jointly optimizing temporal compression and hardware dataflow. Yet, these designs often rely on complex neuron models, offline redundancy analysis, limiting their scalability and real-time applicability. Consequently, despite progress in temporal compression algorithms, an evident mismatch remains between algorithmic efficiency and hardware realizability.

To address these limitations, this work presents STISA, a unified algorithm-hardware co-design framework that achieves low-latency and energy-efficient SNN inference through tight integration of temporal compression and spatiotemporal parallel mapping. On the algorithm side, Temporal Splitter Compression (TSC) achieves competitive accuracies of 96.38\%, 81.10\%, 68.51\%, and 82.40\% on CIFAR-10, CIFAR-100, ImageNet, and DVS-CIFAR10 respectively, while reducing synaptic operations by 51.71\%, 46.71\%, 33.6\%, and 24.44\%. On the hardware side, FPGA prototypes of SCNN9, ResNet18, and ResNet34 achieve speedups of 7.96×, 3.98×, and 7.99×, while reducing power by 17.7\%, 28.1\%, and 30.4\%, and LUT utilization by 20.9\%, 29.8\%, and 30.1\%. These results confirm that STISA provides a scalable, energy-efficient, and real-time deployable solution under limited hardware resources. The proposed framework introduces three synergistic innovations:
\begin{itemize}
\item{
\textbf{Temporal Splitter Compression:} a lightweight temporal compression method with a Temporal Split Structure (TSS), which decouples and redistributes timesteps across parallel branches to suppress temporal redundancy while maintaining accuracy and temporal dynamics.
}
\item{
\textbf{Spatio-Temporal Parallel Architecture (STPA):}
a spatiotemporal hardware mapping strategy combining Temporal-Prioritized Output-Stationary (TPOS) dataflow and Block-Mux Streaming (BMS) architecture. STPA enhances locality of both weight and membrane potentials, eliminates membrane-potential storage, and enables efficient spatiotemporal parallel execution.
}
\item{
\textbf{Unified Co-Optimization Framework:}
TSC and STPA are integrated into a unified algorithm-hardware framework with a bandwidth-aware resource allocation strategy, achieving balanced accuracy, latency, and energy efficiency while flexibly supporting diverse SNN models and FPGA platforms.
}
\end{itemize}

The remainder of this paper is organized as follows. \secref{background} reviews SNN fundamentals, including neuron models, temporal compression in SNNs, dataflow and architectural considerations. \secref{algorithm} introduces the TSC technique for the compute engine and reports its algorithmic results. \secref{hardware_overview} motivates the hardware design and provides an overview. \secref{hardware} details the hardware implementation. \secref{results} presents and compares the hardware evaluation results. Finally, \secref{conclusion} concludes the paper.
\begin{figure*}[tbp]
    \centering
    \vspace{-10pt}
    \includegraphics[width=\linewidth]{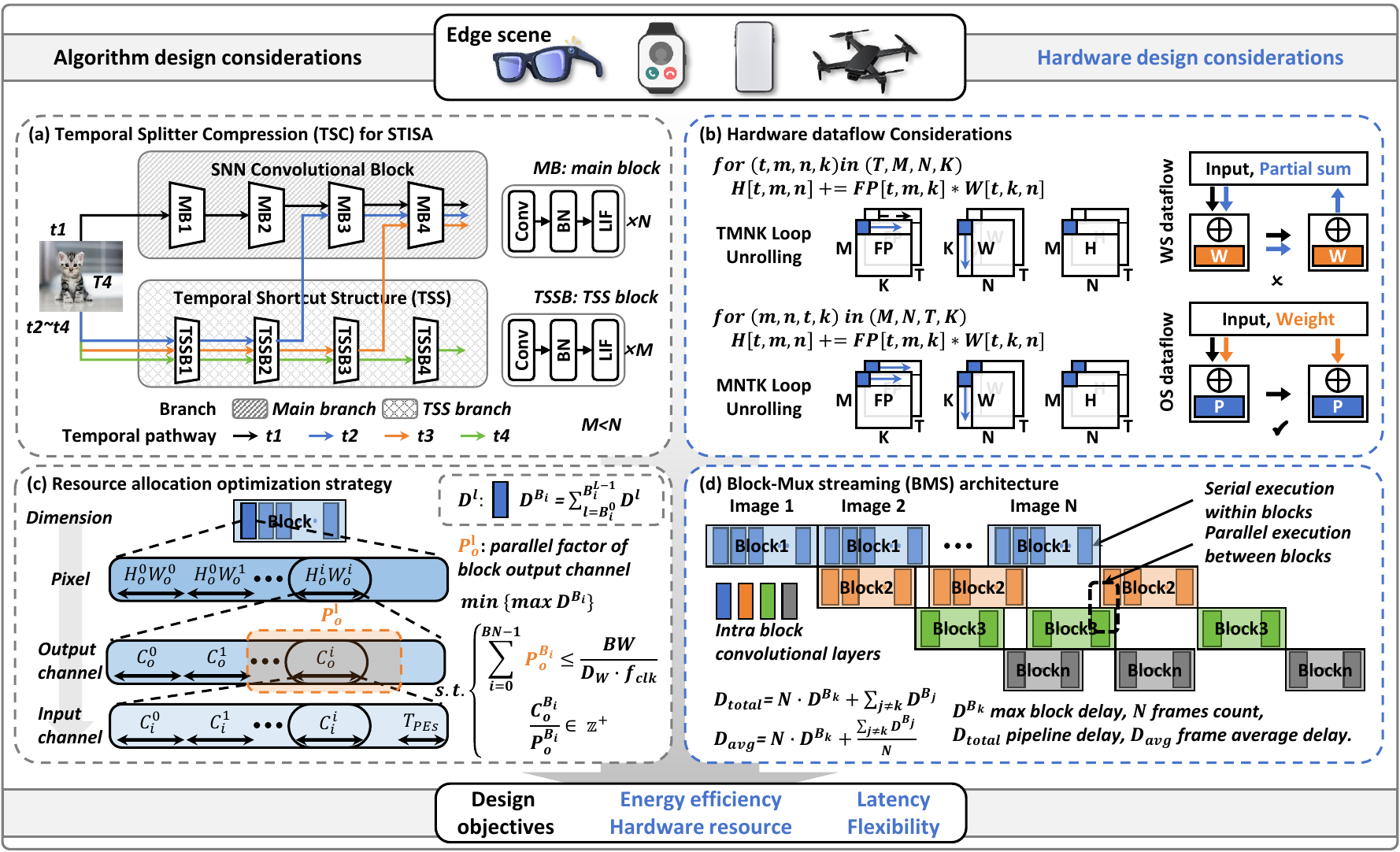}
    \vspace{-10pt}
    \caption{
    Algorithm-hardware co-design of the STISA: (a) TSC enables block-wise adaptive timesteps compression; (b) hardware dataflow comparison; (c) load-aware output-channel parallelism; (d) BMS architecture with cross-block pipelining and intra-block time-division multiplexing.
    }
    \vspace{-20pt}
    \label{fg_ah}
\end{figure*}
\section{Background} \label{background}
\subsection{SNN Neuron Model}
In this work, we adopt the standard leaky integrate-and-fire (LIF) neuron model~\cite{lif}, its membrane potential $v$ evolves over time $t$ according to \eqref{EQ_LIF}, where $I$ denotes the input current, $R$ the resistance, and $\tau_{m}$ the time constant\deleted{of the membrane potential}. Once $v$ reaches the threshold $V_{th}$, the neuron emits a spike and resets to $v_{r}$.
\begin{equation}
\label{EQ_LIF}
\tau_{m} \frac{dv}{dt}=-v+R I(t), \quad v<V_{th}.
\end{equation}

For hardware realization, the discrete-time formulation of the LIF dynamics can be divided into three computational phases:
\paragraph{Input Current Accumulation Phase}
At each discrete timestep, all presynaptic spikes are integrated as \eqref{EQ_LIF_sub1}.
\begin{equation}
\label{EQ_LIF_sub1}
I_{i}[t]=\sum_{j} w_{i j} s_{j}[t]+b_{i}.
\end{equation} 
\paragraph{Membrane Potential Update Phase}
The membrane potential is updated based on the previous state and the accumulated input current, expressed as
\eqref{EQ_LIF_sub2}.
\begin{equation}
\label{EQ_LIF_sub2}
h_{i}[t]=\left(1-\frac{1}{\tau_{m}}\right) v_{i}[t]+I_{i}[t].
\end{equation} 
\paragraph{Output Spike Generation Phase}
When the potential exceeds the threshold, the neuron emits a spike and resets its membrane potential, as \eqref{EQ_LIF_sub3}.
\begin{equation}
\label{EQ_LIF_sub3}
\left(v_{i}[t+1], s_{i}[t+1]\right)=\left\{\begin{array}{l}
\left(h_{i}[t], 0\right), \quad h_{i}[t]<V_{t h} \\
(u_{r},1), \quad h_{i}[t] \geq V_{t h}.
\end{array}\right. 
\end{equation} 

From a hardware perspective, two operations dominate system complexity:
(i) input accumulation, which incurs substantial computational load due to the dense synaptic connectivity; and
(ii) membrane potential update, which requires frequent storage access and incurs significant read-write overhead across timesteps.
 
\subsection{SNNs Temporary Compression} \label{IIB}
Uniformly shrinking the timestep across all layers is an overly aggressive compression strategy. In contrast, layer-wise adaptive timesteps can maintain or even improve accuracy with fewer timesteps on average. With inter-layer temporal demand disparities being objective and quantifiable, layer-wise adaptation provides a closer fit to the intrinsic temporal dynamics of SNNs.

Evidence from temporally adaptive SNNs supports this view. Liu et al. \cite{teas} observed that most directly-trained SNNs adopt synchronized, layer-uniform timesteps; while simple to implement, such designs ignore cross-layer differences in temporal dynamics, producing redundant spikes, repeated computation, and inflated inference latency under mismatched temporal demands. Moreover, naively shortening the global time window discards useful temporal information propagated across layers and leads to substantial accuracy degradation. Information-theoretic analysis further indicates pronounced cross-layer disparity. Using layer-wise entropy, Xie et al. \cite{Xie} reported that activation entropy (in Shannon bits) varies markedly by depth: deeper layers often exhibit higher entropy (0.41-0.72 bit), implying stronger temporal dependency and greater computational demand, whereas other layers show lower entropy (0.01-0.13 bit), indicating redundancy. Under such imbalance, configuring all layers with the same timestep based on the network’s maximum Shannon value inevitably introduces temporal redundancy and unnecessary latency in low-entropy layers.

Shallow layers primarily extract low-level features, and their membrane potentials and firing patterns typically stabilize within fewer timesteps. Deep layers, however, require longer accumulation windows to integrate information over time and form high-level semantics. A uniform timestep therefore imposes a ``short-long'' mismatch, causing redundant spike computations in shallow layers and increasing inference latency.

To avoid information loss from globally shortening timesteps while exploiting inter-layer temporal heterogeneity, this work presents TSC. Unlike prior approaches \cite{teas, Xie}, TSC raises the adaptation granularity to the block level (a block comprising multiple convolutional layers) and performs adaptive timestep allocation at this level (see \fgref{fg_ah} (a)). TSC assigns longer windows to deeper, higher-entropy, or timestep-sensitive blocks, while contracting the windows for shallow blocks with clearer redundancy, achieving a better accuracy-latency-energy trade-off. Implementation details are provided in \secref{algorithm}.

\subsection{TPOS Dataflow and BMS Architecture Hardware Considerations}
To shorten latency and reduce cross timestep memory traffic, we evaluate serial and fully parallel temporal unrolling (\fgref{fg_ah} (b)). Serial inference has a smaller hardware footprint but incurs higher end-to-end latency and frequent off-chip accesses. Fully parallel execution enables single-shot inference, yet its compute and storage grow linearly with the number of timesteps, and redundancy persists. 
To control bandwidth and buffering cost, we further compare output-stationary (OS) and weight-stationary (WS) dataflows and identify the membrane potential (wide bit width arising from channel-wise accumulation) as the dominant storage and bandwidth bottleneck. Based on these tradeoffs, we adopt a TPOS dataflow that prioritizes temporal unrolling, reuses weights across timesteps, and keeps partial sums stationary, thereby eliminating membrane-potential buffering; together with TSC, compressed timesteps are applied within the blocks on the main branch.
At the architectural level, an overlap-based design (a single engine time-multiplexed across layers) is reconfigurable but control-heavy, while a streaming design (per-layer dedicated hardware) simplifies per-layer tuning but scales poorly to deeper models. These tradeoffs motivate the BMS architecture (\fgref{fg_ah} (d)): cross-block pipelining and intra-block time-division multiplexing improve utilization and scalability; block-level homogeneity (aligned feature/weight shapes and timesteps within a block) reduces coordination overhead and resource fragmentation; and intra-block output-channel parallelism (\fgref{fg_ah} (c)) enables load-aware resource allocation and balancing. 
Overall, TSC-TPOS-BMS turns algorithm-level temporal compression into quantifiable gains in latency, energy, and scalability.

\begin{figure}[tbp]
    \centering
    \vspace{-10pt}
    \includegraphics[width=\columnwidth]{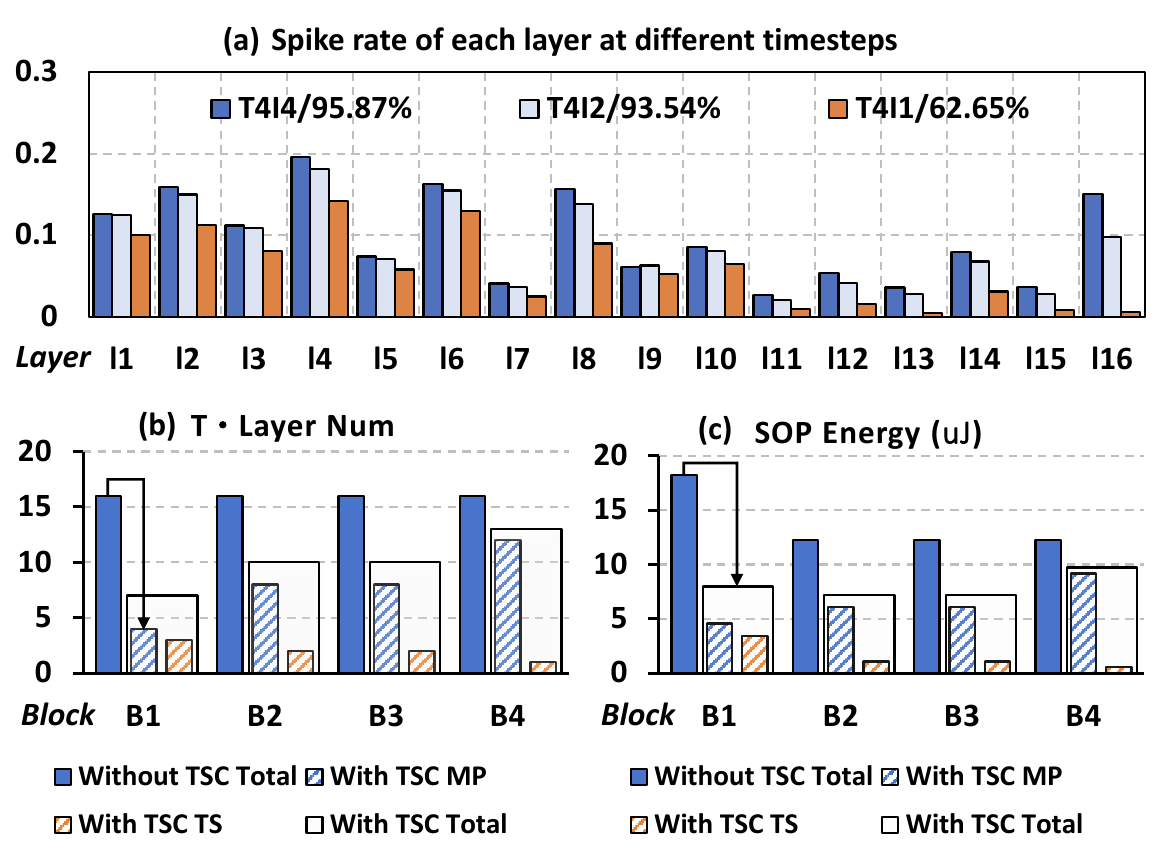}
    \vspace{-10pt}
    \caption{
    Motivation and quantitative analysis of TSC: (a) effect of inference timesteps on spike firing rate and model accuracy (standard direct training, CIFAR-10 on ResNet18); (b) TSC reduces the number of critical paths; (c) theoretical reduction of synaptic-operation energy with TSC.
    }
    \vspace{-20pt}
    \label{fg_tsc_opch}
\end{figure}

\begin{figure*}[tbp]
    \centering
    \vspace{-10pt}
    \includegraphics[width=\linewidth]{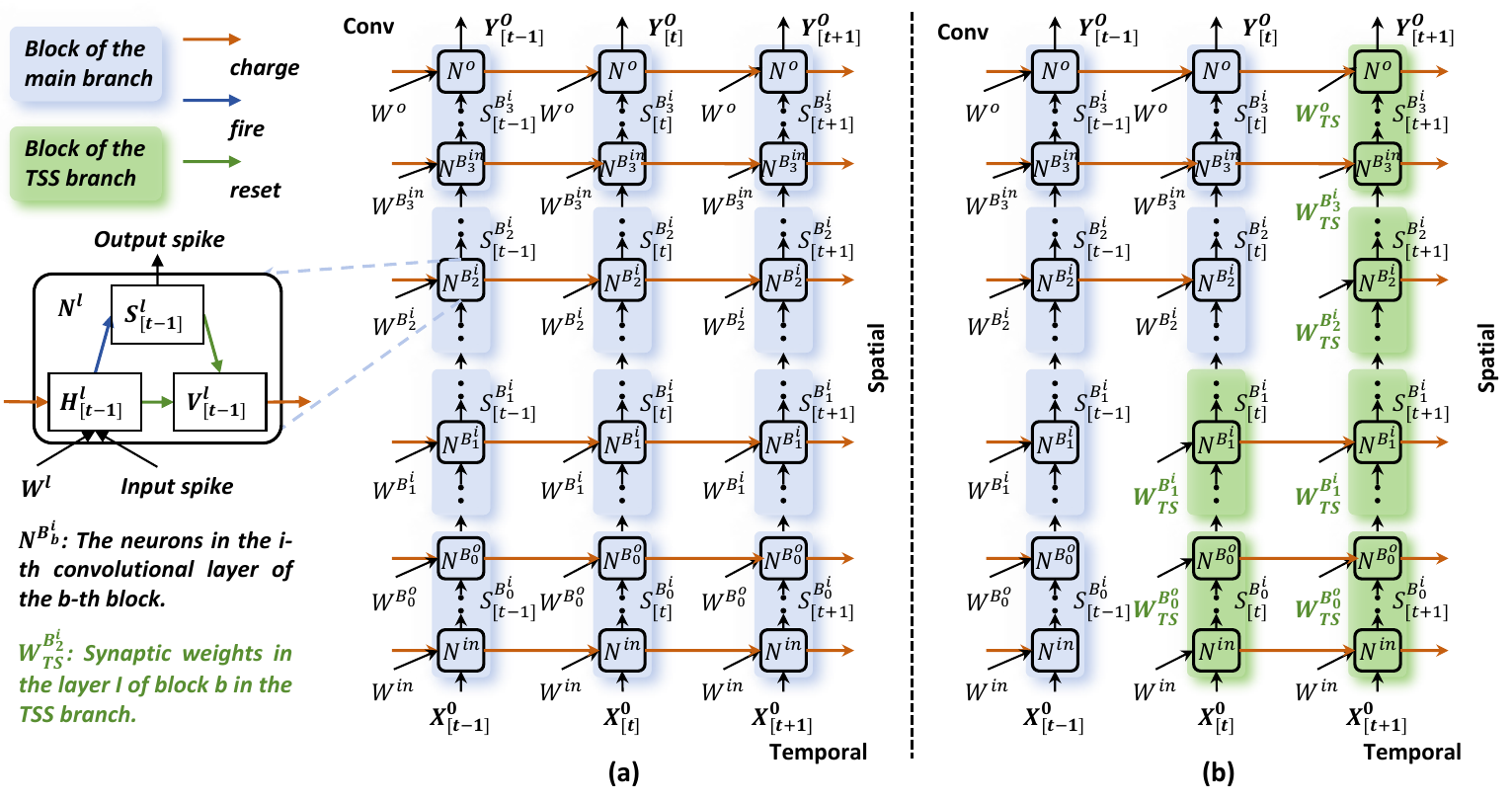}
    \vspace{-20pt}
    \caption{
    Overview of TSC. Shown are (a) the forward-propagation computation graph of a standard SNN and (b) the corresponding graph after TSC is applied.
    }
    \label{fg_tsc_alg}
    \vspace{-20pt}
\end{figure*}

\section{Proposed TSC Algorithm for Single Shot Inference Engine} \label{algorithm}
\subsection{Temporal Splitter Compression Technique} \label{IIIA}
\subsubsection{Comparison Between Standard and TSC-Compressed SNNs}
When the inference timestep is uniformly compressed across all layers, deeper networks exhibit higher sensitivity to temporal reduction, leading to pronounced accuracy degradation and marginal performance loss. In contrast, shallow layers tend to be more robust to shortened timesteps. As illustrated in \fgref{fg_tsc_opch} (a), under an extreme setting where a model trained with four timesteps is directly inferred with only one timestep, a substantial accuracy drop and reduced spiking activity can be observed. This phenomenon arises because postsynaptic neurons only emit spikes when their membrane potential exceeds the firing threshold; thus, a shortened integration window limits potential accumulation, suppressing firing rates and information throughput. This observation is consistent with the consensus discussed in \secref{IIB}.

Motivated by these findings, the TSC method is presented, where temporal computation is allocated on demand. Specifically, TSC enhances the coarse-grained ``uniform timestep'' strategy (\fgref{fg_tsc_alg} (a)) to a fine-grained, block-wise adaptive allocation mechanism (\fgref{fg_tsc_alg} (b)). TSC assigns longer temporal windows to deeper blocks with higher entropy or stronger temporal sensitivity, while allocating shorter windows to shallow blocks that exhibit significant redundancy. This design achieves a better trade-off among accuracy, latency, and energy efficiency.

In implementation, TSC introduces a TSS that diverts part of the timesteps from the main branch to an auxiliary TSS branch. When additional temporal features are required, the main branch retrieves them from the TSS. The depth of TSS can be flexibly configured before training to meet various accuracy-efficiency targets. This structure fundamentally modifies the standard SNN’s forward mode, where all layers are synchronously propagated across all timesteps. Instead, temporal computation is asynchronously and adaptively scheduled between the main and auxiliary branches, improving the precision and controllability of temporal resource allocation. 
For instance, in a spiking ResNet18 model trained with four timesteps, TSC assigns [1, 2, 2, 3] timesteps to its four main blocks. Evaluations show that while the layer-timestep product (LT) and synaptic operation (SOP) energy are significantly reduced (\fgref{fg_tsc_opch} (b), \fgref{fg_tsc_opch} (c)), the accuracy degradation remains below 0.6\%. These results demonstrate that TSC effectively achieves adaptive temporal compression while preserving the intrinsic spatiotemporal dynamics of SNNs.

\subsubsection{Forward Propagation and Loss Function Modeling of TSC} 
The forward propagation of the proposed TSC method is formulated in \eqref{EQ_Y}, corresponding to the computation graph illustrated in \fgref{fg_tsc_alg} (b). Based on the Temporal Efficient Training (TET) method \cite{tet}, the loss function for TSC is defined in \eqref{EQ_Loss}, with its corresponding gradient computation derived in \eqref{EQ_BP}.
During optimization, the cumulative membrane potential $H$ of the output-layer neurons, as depicted in \fgref{fg_tsc_alg}, is directly employed for loss evaluation and backpropagation.
\begin{equation}
\label{EQ_Y}
\begin{aligned}
Y_{TSC}^{B_{b}^{i}}&=Y^{B_{b}^{i}}+Y_{TSS}^{B_{b}^{i}} \\
&=\sum_{t=0}^{t_{b}-1} S^{B_{b}^{i}}(t) \cdot W + \sum_{t=t_{b}}^{T-1} S_{T S S}^{B_{b}^{i}}(t) \cdot W_{TSS}
\end{aligned}
\end{equation}
\begin{equation}
\label{EQ_Loss}
\begin{aligned}
L_{TSC} &= L+L_{TSS} \\
&=\frac{1}{T}\left\{\sum_{t=0}^{t_{b}-1} L_{C E}\left[Y^{o}(t), y\right] + \sum_{t=t_{b}}^{T-1} L_{C E}\left[Y_{T S S}^{o}(t), y\right]\right\}
\end{aligned}
\end{equation}
\begin{equation}
\label{EQ_BP}
\begin{aligned}
\frac{\partial L_{TSC}}{\partial W_{TSC}} 
&= \frac{\partial L}{\partial W} + \frac{\partial L_{TSS}}{\partial W_{TSS}} \\
&= \frac{1}{T} \left\{ \sum_{t=0}^{t_{b}-1} \left[ S_{f}(Y^{o}(t)) - y \right] \cdot \frac{\partial Y^{o}(t)}{\partial W} \right. \\
& \quad + \left. \sum_{t=t_{b}}^{T-1} \left[ S_{f}(Y_{T S S}^{o}(t)) - y \right] \cdot \frac{\partial Y^{o}(t)}{\partial W_{TSS}} \right\}
\end{aligned}
\end{equation}

\subsubsection{Training and Inference Procedure of TSC}
The training of the proposed TSC method follows the standard backpropagation-through-time (BPTT) \cite{stbp} pipeline, while incorporating the TSS auxiliary structure described in \secref{IIIA}. During training, neuron membrane potentials are iteratively updated over discrete timesteps in both the main and TSS branches. 
To ensure stable convergence, the temporal dynamics of the TSS branch are jointly optimized with the main branch, while their parameters remain independent. This cooperative optimization enables the model to learn complementary temporal representations across branches, enhancing both learning efficiency and temporal expressiveness.
During inference, only the forward propagation path is activated. The TSS branch processes a reduced subset of timesteps to provide temporal context for the main branch while mitigating redundant computation. This selective execution effectively reduces SOPs and memory accesses, improving both energy efficiency and throughput.
Compared with conventional SNN inference, the proposed TSC method achieves comparable accuracy with substantially lower computational cost, enabling single-shot inference suitable for low-latency and energy-constrained neuromorphic hardware.

\subsection{TSC Experimental Results and Analysis} \label{IIIB}
The effectiveness of the proposed TSC framework is evaluated on both static and neuromorphic datasets, including CIFAR-10, CIFAR-100 \cite{cifar}, ImageNet \cite{imagenet}, and DVS-CIFAR10 \cite{dvs_cifar10}. Following prior work, spiking ResNet18, ResNet19 and ResNet34 \cite{resnet} are adopted for the CIFAR-series and ImageNet datasets, respectively, while VGGSNN \cite{vgg} is used for DVS-CIFAR10.
\tabref{tab_al_tsc_result} summarizes the comparison between TSC and several representative temporal compression methods, including layer-wise configurable timesteps~\cite{Xie}, dynamic timestep exploration~\cite{Seenn, Iadyt}, and direct-training-based approaches~\cite{Diet,tet,chowdhury}. The baseline networks are trained with 4 timesteps in total.
After applying TSC, the main-branch timestep configurations are set to [1, 2, 2, 3] for CIFAR-10, [1, 2, 3, 3] for CIFAR-100 and ImageNet, leading to accuracy drops of 0.29\%, 1.05\%, and 1.39\%, respectively. For DVS-CIFAR10, configurations [4, 6, 8, 8] and [4, 6, 6, 8] result in accuracy decreases of 0.1\% and 0.6\%, respectively. These results demonstrate that TSC enables flexible temporal compression, allowing a tunable trade-off between compression efficiency and classification accuracy across diverse datasets.

While SEENNI and SEENNII achieve about a 70\% reduction in the number of operations on CIFAR-10 and CIFAR-100, our method attains 0.31\% and 0.87\% higher classification accuracies, respectively, at the expense of a slightly smaller reduction in operations. On the more challenging ImageNet dataset, our approach further improves over SEENNI by 18.1\% in operation reduction and by 3.85\% in accuracy. Overall, on static datasets our method provides a more favorable accuracy-efficiency trade-off than representative existing approaches, and it also delivers competitive recognition performance on dynamic datasets. 
It should be noted that, although sample-dependent dynamic timestep scheduling can effectively reduce the average number of operations per sample, the varying hardware resource demands across samples induce irregular inference patterns and degrade temporal parallelism. This irregularity increases the complexity of hardware implementation, as it hinders deterministic pipeline scheduling and makes it difficult to achieve a better trade-off between efficiency and resource utilization. In contrast, the proposed TSC scheme provides a uniform reduction in computational operations for all input samples, fundamentally alleviating the above issues and enabling more efficient and predictable hardware deployment.

\begin{table*}[!t]
    \renewcommand{\arraystretch}{1.25}
    \small
    \vspace{-10pt}
    \setlength{\tabcolsep}{3pt}
    \centering
    \caption{Comparison with SOTA Temporal Pruning Works}
    \label{tab_al_tsc_result}
    \begin{threeparttable}
    \begin{tabularx}{\linewidth}{l|c|C|C|C|C|C|C}  
    \Xhline{1pt}
     & Dataset & Method & Model & FT\tnote{1} & CT\tnote{1} & RO\tnote{1} & Accuracy \\ 
    \Xhline{0.5pt}
    
    Xie et al.\cite{Xie}                    & CIFAR-10      & LCT       & ResNet18  & 6         & 3.75    & 37.5      & 83.87               \\
    Xie et al.\cite{Xie}                    & CIFAR-10      & LCT       & ResNet18  & /         & /       & /         & 92.14               \\
    Li et al.\cite{Seenn}                   & CIFAR-10      & SEENNI    & ResNet19  & 4         & 1.09    & 72.75     & 96.07               \\
    Li et al.\cite{Seenn}                   & CIFAR-10      & SEENNII   & ResNet19  & 4         & 1.08    & 73        & 96.01               \\
    Li et al.\cite{Iadyt}                   & CIFAR-10      & Iadyt     & ResNet19  & 4         & 1.46    & 63.5      & 93.54               \\
    Rathi et al.\cite{Diet}                 & CIFAR-10      & DietSNN   & ResNet20  & 5         & 5       & /         & 91.78               \\   
    Deng et al.\cite{tet}                   & CIFAR-10      & TET       & ResNet19  & 2         & 2       & /         & 94.16               \\
    Chowdhury et al.\cite{chowdhury}        & CIFAR-10      & IIRSNN    & VGG16     & 1         & 1       & /         & 93.05               \\
    \textbf{This work}                      & CIFAR-10      & STISA     & ResNet18  & 4         & 4       & /         & $95.87\pm0.14$      \\
    \textbf{This work}                      & CIFAR-10      & STISA     & ResNet18  & 4         & 2       & 40.5      & $95.29\pm0.16$      \\
    \textbf{This work}                      & CIFAR-10      & STISA     & ResNet19  & 4         & 4       & /         & $96.67\pm0.11$      \\
    \textbf{This work}                      & CIFAR-10      & STISA     & ResNet19  & 4         & 1.25\tnote{2}    & 51.71     & $96.38\pm0.13$      \\
    
    \Xhline{0.5pt}           
    Li et al.\cite{Seenn}                   & CIFAR-100     & SEENNI    & ResNet19  & 4         & 1.19    & 70.25     & 79.56                  \\
    Li et al.\cite{Seenn}                   & CIFAR-100     & SEENNII   & ResNet19  & 4         & 1.21    & 69.75     & 80.23                  \\  
    Li et al.\cite{Iadyt}                   & CIFAR-100     & Iadyt     & ResNet19  & 4         & 1.9     & 52.5      & 73.48                  \\
    Rathi et al.\cite{Diet}                 & CIFAR-100     & DietSNN   & ResNet20  & 5         & 5       & /         & 64.07                  \\  
    Deng et al.\cite{tet}                   & CIFAR-100     & TET       & ResNet19  & 2         & 2       & /         & 72.87                  \\
    Chowdhury et al.~\cite{chowdhury}       & CIFAR-100     & IIRSNN    & VGG16     & 1         & 1       & /         & 70.15                  \\
    \textbf{This work}                      & CIFAR-100     & STISA     & ResNet18  & 4         & 4       & /         & $78.46\pm0.15$         \\
    \textbf{This work}                      & CIFAR-100     & STISA     & ResNet18  & 4         & 2.25\tnote{2}    & 41.6      & $78.15\pm0.17$         \\
    \textbf{This work}                      & CIFAR-100     & STISA     & ResNet19  & 4         & 4       & /         & $82.15\pm0.11$         \\
    \textbf{This work}                      & CIFAR-100     & STISA     & ResNet19  & 4         & 1.5\tnote{2}     & 46.71     & $81.10\pm0.14$         \\

    \Xhline{0.5pt}          
    Li et al.\cite{Seenn}                   & ImageNet      & SEENNI    & ResNet34  & 4         & 3.38    & 15.5      & 64.66                   \\
    Li et al.\cite{Seenn}                   & ImageNet      & SEENNII   & ResNet34  & 4         & 2.40    & 40        & 64.18                   \\
    Rathi et al.\cite{Diet}                 & ImageNet      & DietSNN   & VGG16     & 5         & 5       & /         & 69.00                   \\ 
    Deng et al.\cite{tet}                   & ImageNet      & TET       & ResNet34  & 4         & 4       & /         & 68.00                   \\
    Chowdhury et al.~\cite{chowdhury}       & ImageNet      & IIRSNN    & VGG16     & 1         & 1       & /         & 69.00                   \\
    \textbf{This work}                      & ImageNet      & STISA     & ResNet34  & 4         & 4       & /         & $69.90\pm0.11$          \\
    \textbf{This work}                      & ImageNet      & STISA     & ResNet34  & 4         & 2.25\tnote{2}    & 33.6      & $68.51\pm0.16$          \\
    
    \Xhline{0.5pt}      
    Li et al.\cite{Seenn}                   & DVS-CIFAR10   & SEENNI    & VGGSNN    & 10        & 5.17    & 48.3      & 82.7\tnote{3}           \\
    Li et al.\cite{Seenn}                   & DVS-CIFAR10   & SEENNII   & VGGSNN    & 10        & 4.49    & 55.1      & 82.6\tnote{3}           \\
    Li et al.\cite{Iadyt}                   & DVS-CIFAR10   & Iadyt     & VGG16     & 10        & 5.25    & 47.5      & 74.40                   \\
    Deng et al.\cite{tet}                   & DVS-CIFAR10   & TET       & VGGSNN    & 10        & 10      & /         & 77.33                   \\
    Deng et al.\cite{tet}                   & DVS-CIFAR10   & TET2      & VGGSNN    & 10        & 10      & /         & 83.17\tnote{3}          \\
    \textbf{This work}                      & DVS-CIFAR10   & STISA     & VGGSNN    & 10        & 10      & /         & $82.50\pm0.07$\tnote{3}       \\
    \textbf{This work}                      & DVS-CIFAR10   & STISA     & VGGSNN    & 10        & 6.5\tnote{2}     & 24.44     & $82.40\pm0.09$\tnote{3}       \\  
    \textbf{This work}                      & DVS-CIFAR10   & STISA     & VGGSNN    & 10        & 6\tnote{2}       & 31.42     & $81.90\pm0.11$\tnote{3}       \\  
    \Xhline{1pt}
    \end{tabularx}

    \label{tab:gops1}
    \begin{tablenotes}
        \item[1] FT: Full timesteps, CT: Compressed timesteps, RO: reduced operations (\%).
        \item[2] CT is computed as the average timesteps of all main branch blocks in STISA.
        \item[3] Adopting data augmentation\cite{dataaug}.
    \end{tablenotes}
    \end{threeparttable}
    \vspace{-15pt}
\end{table*}

When temporal serial processing is maintained, TSC can reduce inference latency without increasing hardware cost. Furthermore, introducing temporal parallelism can further reduce latency to the level of a single timestep. Although TSC effectively reduces both latency and power consumption, it introduces two major challenges. 
(i) Load imbalance between the main and auxiliary branches limits system efficiency. The computational load imbalance among branch blocks leads to significant variations in computational cost and latency. Consequently, the end-to-end latency is dominated by the main branch, resulting in degraded system efficiency and reduced hardware utilization of TSS blocks. 
(ii) Block-level adaptive timesteps complicate control and induce resource fragmentation. The core issue arises from idle waiting when the auxiliary branch finishes earlier than the main branch. When the timestep configurations of main and auxiliary branches are inconsistent across blocks (e.g., MB1-T1 with TSSB1-T3 versus MB4-T3 with TSSB4-T1), the idle ratio fluctuates across blocks, leading to inconsistent parallelism and increased scheduling overhead.

\begin{figure*}[tbp]
    \centering
    \vspace{-10pt}
    \includegraphics[width=0.95\linewidth]{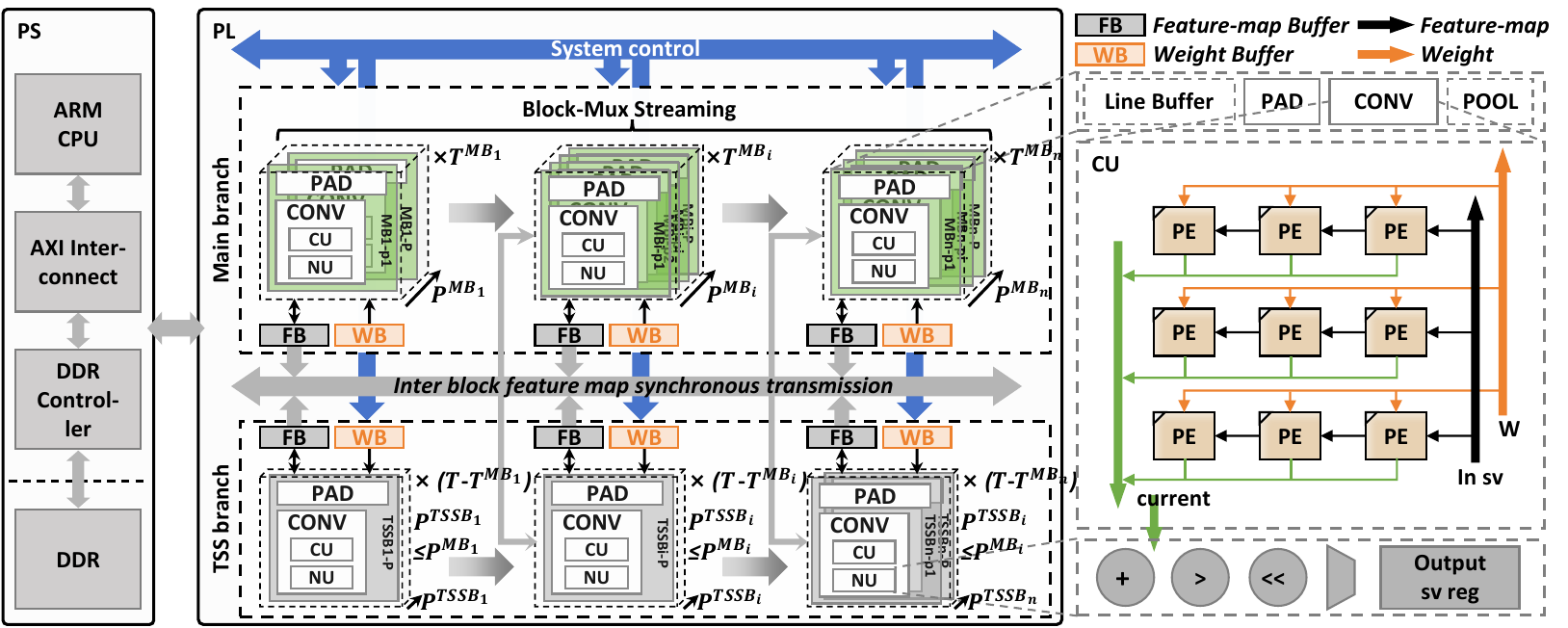}
    \caption{
    Overall STISA hardware architecture (with TSC and spatio-temporal parallelism).
    }
    \label{fg:sti-snn}
    \vspace{-20pt}
\end{figure*}

\section{STISA Hardware overview} \label{hardware_overview}
STISA targets low latency and high energy efficiency for SNN inference via tight algorithm-hardware co-design, while maintaining competitive accuracy. However, algorithmic optimization alone, such as TSC, which suppresses temporal redundancy and reduces computational cost and average latency, cannot achieve single-shot inference (i.e., inference completed within a single forward propagation) and moreover, gives rise to two hardware-level challenges (see \secref{IIIB}). Consequently, appropriate dataflow and architecture strategies are required to translate algorithmic improvements into hardware benefits. This section outlines the hardware design principles and system organization of STISA that address these challenges and complete the co-design.

\textbf{Why the TPOS dataflow?} 
Single-shot inference requires full temporal parallelization. Although existing fully temporal-parallel designs mitigate inter-timestep state traffic, spatial membrane states still incur frequent on/off-chip accesses, which limits latency and energy efficiency. The proposed TPOS (\secref{VA}) (i) prioritizes temporal-loop unfolding to enable parallel input-current accumulation with improved weight locality, and (ii) eliminates membrane-potential buffering and access throughout inference, thereby reducing bandwidth and memory burden while simplifying control. TPOS thus directly addresses the core bottlenecks of latency and state-access suppression.

\textbf{Why the BMS architecture?}  
Under TPOS, the two TSC-related challenges (\secref{IIIB}) remain. BMS (\secref{VB}) addresses them as follows: (i) Block-granular orchestration. Departing from conventional overlap-based or layer-wise streaming architectures, BMS performs block-wise streaming and resource partitioning, which lowers inter-block coordination and scheduling overhead, and alleviates resource fragmentation via homogeneous per-block configurations (identical timestep, feature-map and weight sizes). Time-multiplexed compute and buffers within each block enable deeper networks without additional hardware. The resulting block pipeline with intra-block reuse is well suited to hierarchical networks (e.g., ResNet, VGG), offering generality and scalability. (ii) Load-aware parallelization. BMS employs output-channel parallelization within each block and prioritizes computational resources for latency-critical main-branch blocks. This narrows the latency gap within each main-auxiliary pair, balances load across pairs, and shortens the critical path, thereby improving overall utilization and throughput.

STISA integrates the TPOS dataflow and the BMS architecture to enable temporal-spatio parallel execution, and augments them with TSC’s block-wise adaptive timestep allocation, forming a unified TSC-TPOS-BMS co-optimization framework. TSC reduces the need to linearly scale compute and memory with the temporal-spatial parallelism (TSP) factor. Jointly, these components deliver low latency, high energy efficiency, and scalability under limited bandwidth and hardware resources. Recognizing that different networks and datasets differ in bandwidth and parallelism demands, a bandwidth-aware resource allocation schemeis incorporated. By modeling per-block latency and parallelism under bandwidth constraints, STISA solves a constrained optimization to maximize throughput and equalize per-block latency within a given bandwidth budget, thereby sustaining consistent performance across diverse models and platforms.

\fgref{fg:sti-snn} illustrates an architecture with $n$ main-auxiliary block pairs. A per-block configurable temporal-spatial parallelism is supported, and inter-block feature-map transfers are synchronized to maintain pipeline consistency. Each block integrates dedicated control logic to realize layer-wise time-division multiplexing and to accommodate feature maps with diverse spatial dimensions and channel counts. Private on-chip feature and weight buffers are allocated per block to improve locality. For the execution of a single block under ``one timestep and one output channel'', the convolution module comprises a Compute Unit (CU) and a Neuron Unit (NU). The CU employs systolic array (SA) to perform parallel accumulation of input currents; input features and weights are fetched from the block-local feature and weight buffers, respectively. The NU receives the CU output, compares it against the membrane-potential threshold, updates the neuron state, and determines spike emission. The resulting spike vector is written back to the block’s feature buffer and subsequently consumed by the line buffer of the next convolutional layer.

\begin{figure*}[tbp]
    \centering
    \vspace{-10pt}
    \includegraphics[width=0.95\linewidth]{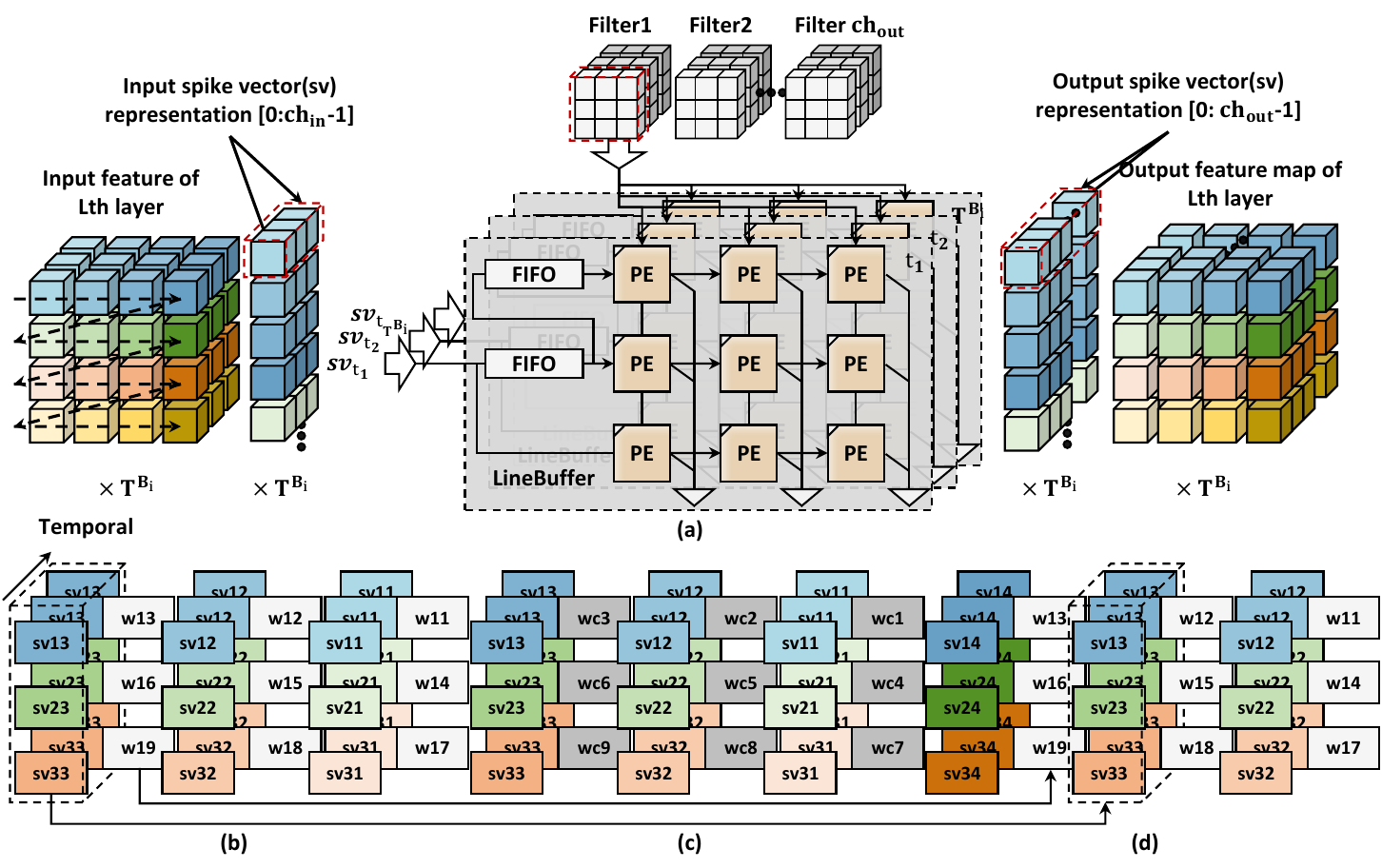}
    \caption{
    (a) TPOS dataflow for convolution layers in STISA and the channel-packed spike vector representation, (b-d) Updates of the input receptive field and weight data in the TPOS dataflow.
    }
    \label{fg:os}
    \vspace{-10pt}
\end{figure*}

\section{Hardware Implementation} \label{hardware}
\subsection{TPOS Dataflow} \label{VA}
\subsubsection{TPOS Dataflow of the Convolutional Layer}
\deleted{
As discussed in \secref{background}, conventional temporal serial processing introduces longer inference latency, frequent external memory accesses for weights, and low bandwidth utilization. Moreover, it requires on-chip storage of neuron states, such as membrane potentials, across timesteps. On the other hand, fully parallelizing all timesteps would lead to excessive hardware costs and redundant computations. Benefiting from the proposed TSC algorithm, each block in the main branch can operate with a compressed timestep. 
}
The TPOS dataflow is shown in \fgref{fg:os}. For the $i$-th main-branch block, the compressed timestep is denoted as $T^{B_i}$. As illustrated in \fgref{fg:os} (a), the CU, composed of multiple PEs, is configured according to $T^{B_i}$ and the convolutional kernel size. PEs within the same row are capable of transmitting input spike vectors to their neighboring PEs, enabling efficient spatial data reuse.
As shown in \fgref{fg:os} (b), the input feature map is convolved with multiple filters, requiring accumulation across input channels. When the input spike vectors are propagated to the rightmost PE, the compute array concurrently processes the spike vectors corresponding to the first receptive field. For each filter, the weights of different channels are sequentially accessed and broadcast to all PEs, allowing temporal weight reuse across timesteps. Each PE accumulates the weighted inputs into its internal partial sum register, as shown in \fgref{fg:os} (c), where $w_{ck}$ denotes the $k$-th weight in the $c$-th channel ($k=1 \sim K_w \times K_h$, $c=1 \sim C_i \times C_o$).
After the CU completes the accumulation for all $C_i$ input channels, each PE outputs its partial sum and immediately begins the next accumulation cycle. The NU subsequently aggregates the CU outputs to obtain the membrane potential $h$, and serially compares it with the firing threshold across timesteps to update neuron states. During this process, spikes are generated and neuron states are refreshed without retaining membrane potentials between timesteps.
Once the CU finishes processing all weights of the $C_o$-th filter, the NU produces the corresponding output spike vector. The CU then updates its receptive field, as illustrated in \fgref{fg:os} (d). Owing to the overlapped pipeline between the CU and NU, the runtime of the NU is effectively hidden under that of the CU. Consequently, no on-chip buffering of neuron state information is required, resulting in improved computation efficiency and reduced memory overhead.
\begin{figure}[tbp]
    \centering
    \vspace{-10pt}
    \includegraphics[width=\columnwidth]{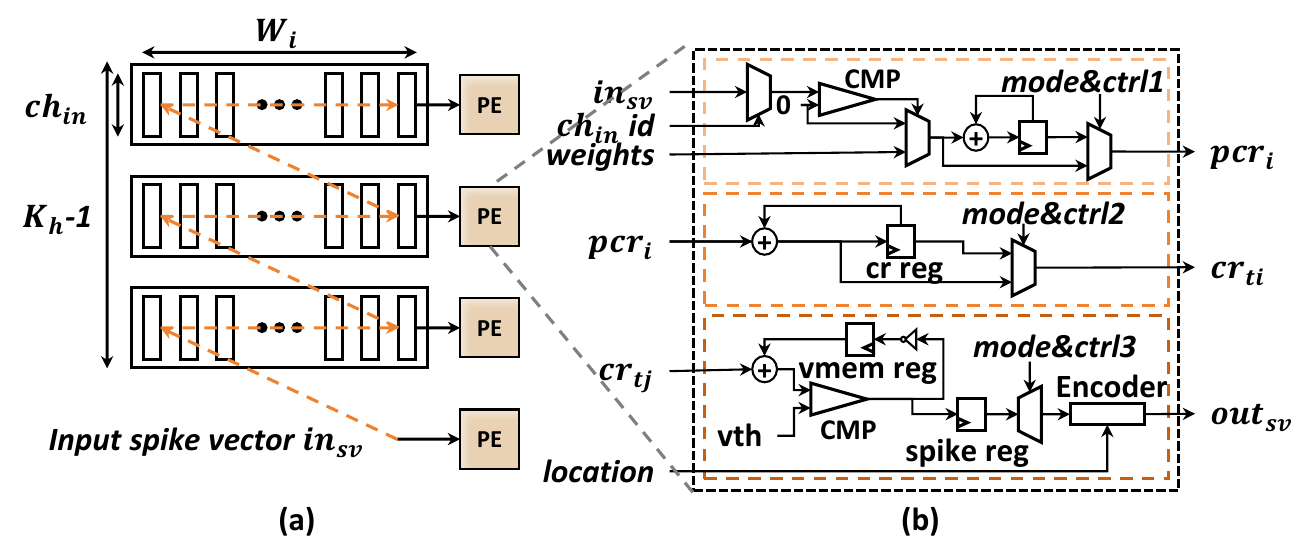}
    \vspace{-10pt}
    \caption{
    Microarchitecture of (a) the line buffer and (b) the processing element.
    }
    \label{fg:linebuffer}
    \vspace{-10pt}
\end{figure} 
\subsubsection{Channel-Packed Spike Vector Representation and Tailored Line-Buffer Microarchitecture} \label{hardware_sv}
As illustrated in \fgref{fg:os} (a), spikes across all channels at the same spatial coordinate are aggregated into a channel-ordered vector, enabling a single memory transaction to fetch multiple channels and thereby improving bandwidth utilization. To further promote on-chip reuse, selected input spike vectors are staged in a line buffer (\fgref{fg:linebuffer} (a)) built from tail-to-head-connected FIFO rows; each FIFO row feeds the aligned PE row in parallel to sustain continuous streaming. For standard convolution, the FIFO count equals the kernel height $K_h - 1$, and each FIFO is sized to a minimum depth of $W_i$ and a bitwidth of $C_i$, matching the length of one packed spike vector.



\subsubsection{Microarchitecture of the Processing Element}
As illustrated in \fgref{fg:linebuffer} (b), the PE comprises three vertically stacked submodules: the channel-wise current accumulator, the pixel-wise current accumulator, and the spike-vector generator. These submodules jointly support standard, depthwise, and pointwise convolution modes, selected by configuration signals \cite{stisnnv1}.

\subsection{Block-Mux Streaming Architecture} \label{VB}
\subsubsection{BMS Architecture Design}
As illustrated in \fgref{fg:sti-snn}, STISA is organized as a block-wise pipeline. To reduce global control overhead and deployment complexity while mitigating resource fragmentation, the convolution compute units and local feature buffers are parameterized and configured on a per-block basis. The timestep allocation for the main-auxiliary block pairs follows the algorithmic in \secref{IIIB}. Given the mismatch in compute intensity and processing rate across blocks, a request-response handshaking scheme is employed to synchronize inter-block feature-map transfers and to maintain a steady pipeline under rate imbalance.

\begin{figure}[tbp]
    \centering
    \vspace{-10pt}
    \includegraphics[width=\linewidth]{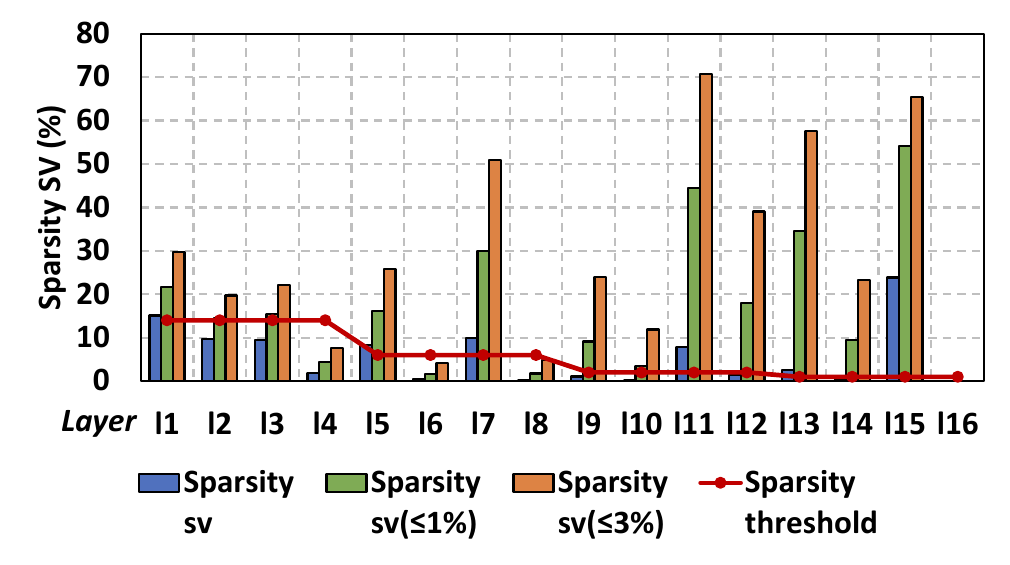}
    \caption{
    Sparsity thresholds and measured sparsity of COO-encoded spike vectors (CIFAR-10 on ResNet18).
    }
    \label{fg_sv_sparsity}
    \vspace{-10pt}
\end{figure}

To reduce inter-block traffic and off-chip accesses, we transmit only non-zero spike vectors and encode them in coordinate (COO) format. According to \eqref{EQ_COO}, sparse transmission becomes bandwidth-efficient once the spike-vector sparsity of the first/last blocks exceeds 38\%/9\% (MNIST), 14\%/1\% (CIFAR-10), and 16\%/1\% (ImageNet) This threshold relaxes as depth increases. Deeper blocks use fewer index bits (logarithmic in feature-map size), while the value bits grow with channel count and dominate; hence the relative gain of sparse encoding increases with depth, even one avoided transmission saves bandwidth. As shown in \fgref{fg_sv_sparsity}, with sparse encoding alone, the first convolutional layer exhibits 15.14\% spike-vector sparsity (exceeding the 14\% CIFAR-10 threshold), yielding a 1.72\% reduction in off-chip memory traffic. With pruning (i.e., skipping transmission of a spike vector when its non-zero ratio is below 1\%), off-chip traffic drops by 9.8\% and the system-level memory traffic by 4.87\%. Relaxing the pruning threshold to 3\% further reduces off-chip traffic by 19.06\% and system traffic by 15.46\%.
\begin{equation}
\label{EQ_COO}
\left[C_{o}+f\left(H_{o}\right)+f\left(W_{o}\right)\right] *(\text{$1-Sparsity$}) \leq C_{o}
\end{equation}

By combining per-block parameterization, inter-block handshaking, and COO-based sparse spike encoding, the proposed block-wise pipelined architecture jointly improves hardware-resource management and data-transfer efficiency.

\subsubsection{Intra-Block Parallel Processing Strategy and Optimization}
\begin{figure}[!tbp]
    \centering
    \includegraphics[width=\linewidth]{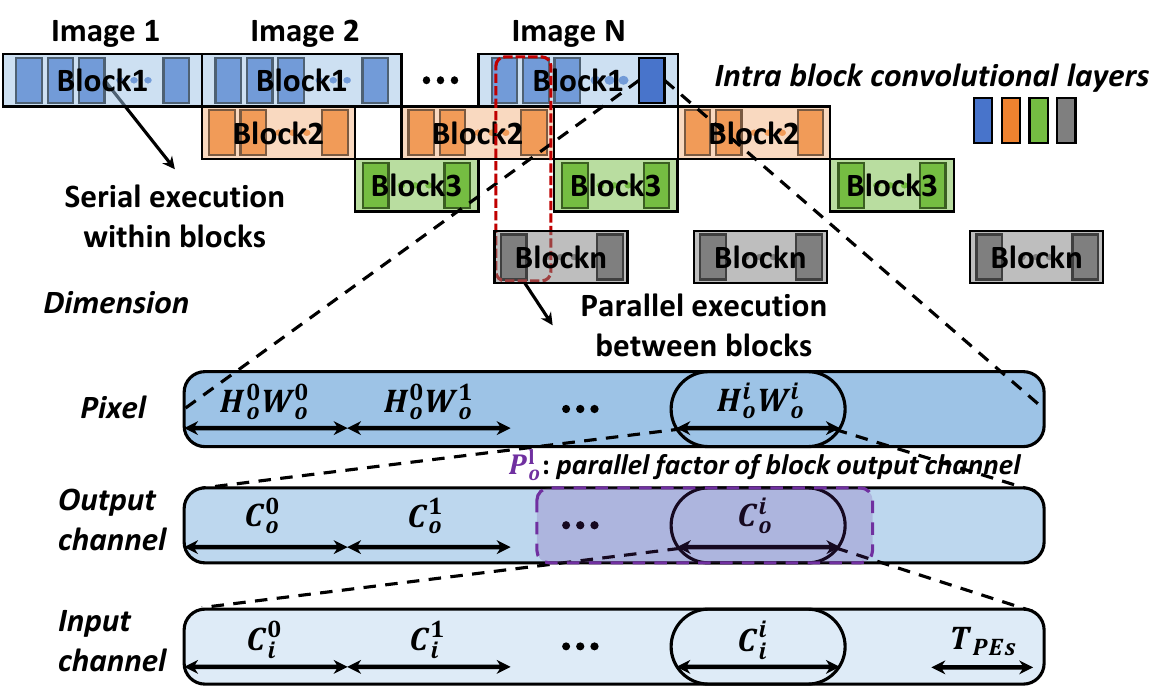}
    \caption{
    BMS inference over $N$ inputs and latency decomposition of intra-block convolution layers.
    }
    \label{fg:pipl}
    \vspace{-10pt}
\end{figure}

\fgref{fg:pipl} illustrates inference of the BMS over $N$ inputs. To enable intra-block output-channel parallelism, we first model the end-to-end pipeline latency and then decompose the latency of a single block with similar layers. Let $D^{B_k}$ denote the maximum latency at the $k$-th convolution block and $N$ the number of processed frames. The total pipeline latency $D_{total}$ follows \eqref{EQ_Ttotal}, and the average per-frame latency $D_{avg}$ follows \eqref{EQ_Tavg}. From \eqref{EQ_Tavg}, the overall throughput is bounded by the slowest convolution block; as $N$ increases, the average latency approaches $D^{B_k}$, and the pipeline latency is amortized over $N$ inputs. To further improve system efficiency, the latency of $i_{th}$ block, $D^{B_i}$, is decomposed as shown in \fgref{fg:pipl} and \eqref{EQ_Tconv}. A block-specific output-channel parallel factor $P_{o}^l$ is introduced to tune layer latency $D^l$. 
\begin{equation}
\label{EQ_Ttotal}
D_{total}= N \times D^{B_k}+\sum_{j \neq k} D^{B_j}
\end{equation} 
\begin{equation}
\label{EQ_Tavg}
D_{avg}=D^{B_k}+\frac{\sum_{j \neq k} D^{B_j}}{N}
\end{equation}
\begin{equation}
\label{EQ_Tconv}
D^{B_{i}}=\sum_{l=B_{i}^{0}}^{B_{i}^{L-1}} D^{l} \\
=\sum_{l=B_{i}^{0}}^{B_{i}^{L-1}} H_{o}^{l} \cdot W_{o}^{l} \cdot \frac{C_{o}^{l}}{P_{o}^{B_{i}}} \cdot\left(C_{i}^{l}+N_{pe}^{l}\right) \cdot T_{clk}
\end{equation}

Across different SNNs, the optimization objective is consistent \eqref{EQ_Target}: determine per-block optimal parallel factor to maximize inference performance. Subject to \eqref{EQ_Pconst}, bandwidth requirements are evaluated to ensure the chosen $P_{o}^l$ does not exceed available bandwidth, thereby preserving efficient dataflow. For control simplicity and regular mapping, the parallel factor $P_{o}^l$ is additionally required to be divisible by the number of output channels. The resulting per-block parallel factors are derived under a principle of equivalent optimization, which adaptively balances compute efficiency and bandwidth limits along the throughput-critical path, yielding a scalable solution across diverse network architectures.
\begin{equation}
\label{EQ_Target}
\min \left\{\max D^{B_{i}}\right\}\\
\end{equation}
\begin{equation}
\label{EQ_Pconst}
s.t.\left\{\begin{matrix} 
 \sum_{i=0}^{BN-1} P_{o}^{B_{i}} \leq \frac{BW}{D_{W} \cdot f_{clk}} \\
 \frac{C_{o}^{B_{i}}}{P_{o}^{B_{i}}} \in \mathbb{Z}^{+}
\end{matrix}\right. 
\end{equation}

\begin{figure}[tbp]
    \centering
    \includegraphics[width=0.95\columnwidth]{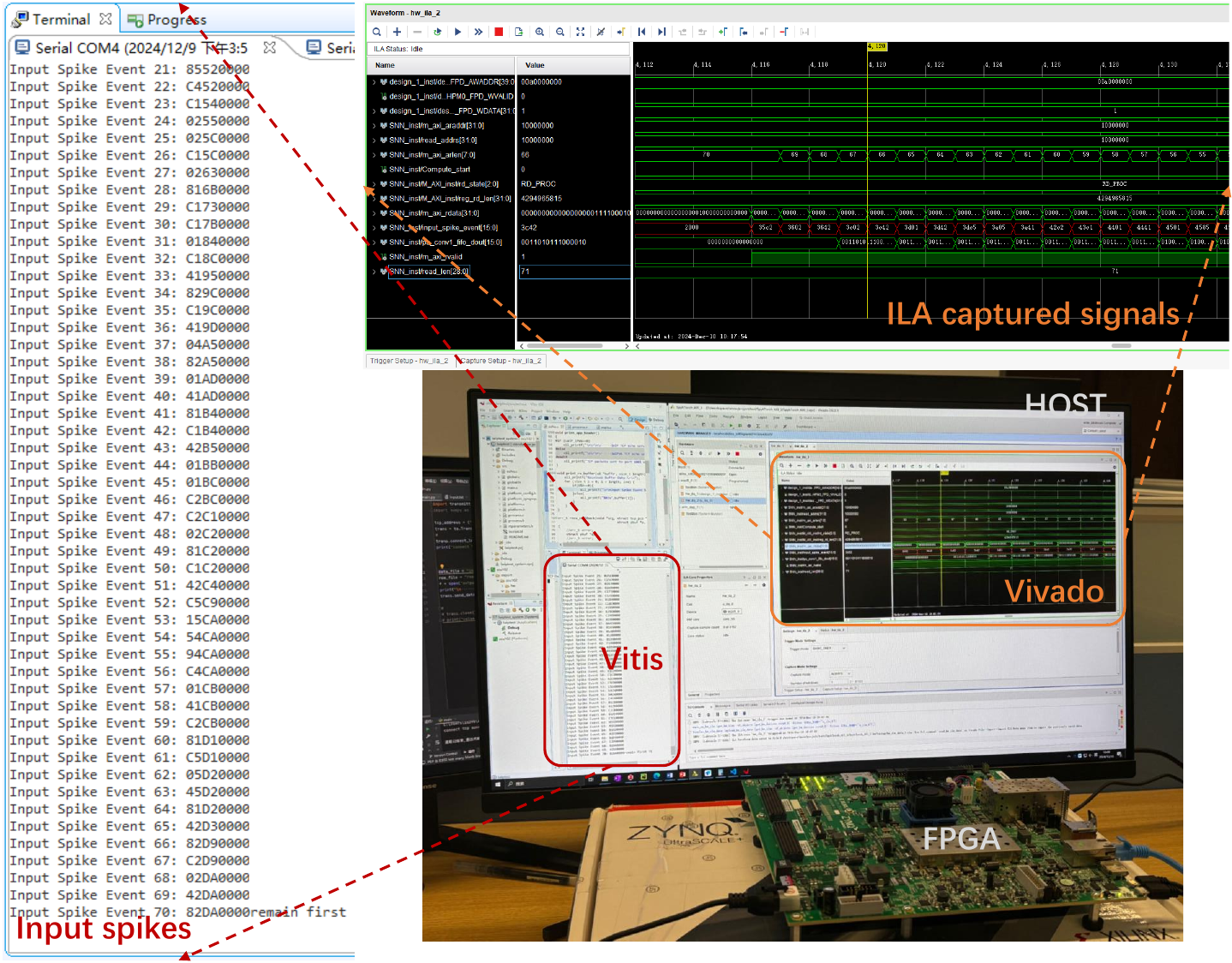}
    \caption{
     Real-world test setup of a classification system for functional validation of the proposed STISA.
    }
    \label{fg_hostandfpga}
    \vspace{-10pt}
\end{figure}

\section{Experimental Setup and Hardware Evaluation}  \label{results}
\subsection{Experimental Setup}
To validate the proposed STISA, we evaluate its performance on multiple SNN models and datasets. For image classification, SCNN9 and spiking ResNet18 are adopted on CIFAR-10, while spiking ResNet34 is used for ImageNet. In all cases, the first convolutional layer performs spike encoding, and the resulting spike trains are fed into the accelerator.
The experimental platform consists of a host-FPGA system, as illustrated in \fgref{fg_hostandfpga}. The Xilinx ZCU102 board with an XCZU9EG FPGA serves as the primary platform. The design is implemented in Verilog HDL and synthesized using Xilinx Vivado 2022.1 at 200 MHz. Vitis 2022.1 is employed to establish a TCP-based interface that transfers input data and control commands between the host and FPGA. All models employ 8-bit fixed-point quantization for weights to ensure fair comparison.

\begin{figure*}[tbp]
    \centering
    \vspace{-10pt}
    \includegraphics[width=0.95\linewidth]{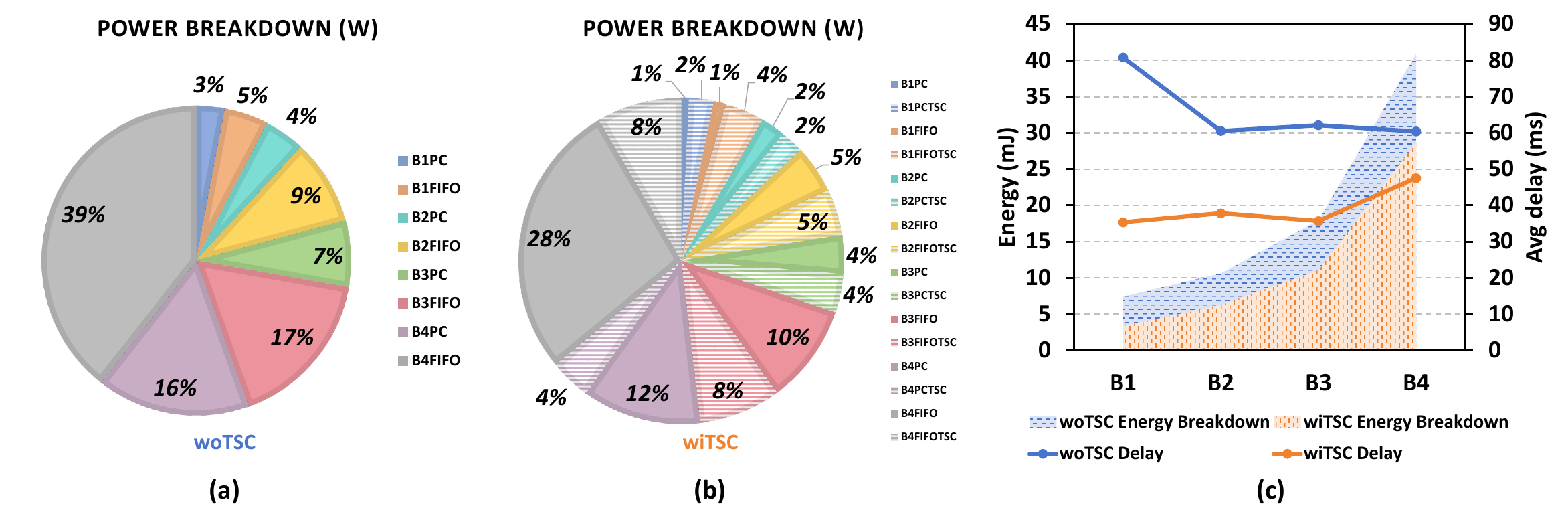}
    \caption{
    Comparison of hardware power distribution, average inference latency, and energy under the TPOS dataflow (without intra-block parallelism, CIFAR-10 on ResNet18), in both TSC-enabled and TSC-disabled configurations. 
    }
    \label{fg_onlyTSC}
    \vspace{-10pt}
\end{figure*}

\subsection{Hardware Evaluation}
\subsubsection{Evaluation of TPOS With and Without TSC}
This subsection investigates the impact of TSC on the TPOS dataflow using the CIFAR-10 dataset with a spiking ResNet18 architecture. In the baseline configuration, the total number of timesteps is 4, where inter-block pipelining is applied but intra-block operations are executed sequentially. After applying TSC, the timestep allocations for the main-branch blocks are set to [1, 2, 2, 3], and the auxiliary branch configurations are summarized in \tabref{tab_snn_models}.
Since all layers within each block share the same hardware resources, the overall resource utilization and power distribution remain approximately the same with or without TSC, as shown in \fgref{fg_onlyTSC} (a), \fgref{fg_onlyTSC} (b). However, the auxiliary branches introduced by TSC exhibit shorter latency and partially offload power consumption from the main branch. Consequently, the energy consumption of the auxiliary blocks decreases, leading to an overall system-level energy reduction. As illustrated in \fgref{fg_onlyTSC} (c), the average latency of main and auxiliary branch is reduced, and per-block energy consumption is improved. Although the integration of TSC within TPOS effectively reduces energy consumption, the system performance is still limited by the block with the maximum latency, which becomes the dominant bottleneck. This imbalance further degrades the hardware utilization of other blocks. We further discuss this issue in the next subsection.

\begin{table}[tbp]
\renewcommand{\arraystretch}{1.25}
\centering
\newcolumntype{N}[1]{>{\raggedright\arraybackslash}p{#1}}
\caption{Standard SNN architecture and its TSC-augmented variant.}
\begin{tabularx}{\columnwidth}{N{2cm} | N{6cm}}
\Xhline{1pt}
SNN Network & Network Structure \\
\Xhline{0.5pt}
SCNN9 & 32×32 [64c3-p2]-[64c3-(128c3)*3-p2]-[128c3-256c3-p2]-[256c3-512c3-p2]-2048-10 \\
\Xhline{0.5pt}
SCNN9-TSC & 32×32 [64c3-p2]-[64c3-(128c3)*1-p2]-[128c3-256c3-p2]-[256c3-512c3-p2] \\
\Xhline{0.5pt}
ResNet18 & 32×32 [(64c3)*4-p2]-[64c3-(128c3)*3-p2]-[128c3-(256c3)*3-p2]-[256c3-(512c3)*3-p2]-AA-512-100 \\
\Xhline{0.5pt}
ResNet18-TSC & 32×32 [(64c3)*1-p2]-[64c3-(128c3)*1-p2]-[128c3-(256c3)*1-p2]-[256c3-(512c3)*1-p2] \\
\Xhline{0.5pt}
ResNet34 & 56×56 [(64c3)*6-p2]-[64c3-(128c3)*7-p2]-[128c3-(256c3)*11-p2]-[256c3-(512c3)*5-p2]-AA-512-1000 \\
\Xhline{0.5pt}
ResNet34-TSC & 56×56 [(64c3)*1-p2]-[64c3-(128c3)*2-p2]-[128c3-(256c3)*2-p2]-[256c3-(512c3)*2-p2] \\
\Xhline{1pt}
\end{tabularx}
\label{tab_snn_models}
\vspace{-10pt}
\end{table}

\subsubsection{Joint Evaluation of TPOS+TSC With and Without BMS}
This subsection evaluates the combined impact of TPOS and BMS architectures, with and without intra-block parallelism, under the TSC configuration. Experiments are conducted on CIFAR-10 and ImageNet datasets using the SCNN9 and ResNet34 networks, respectively. For SCNN9, the timestep and intra-block parallelism of the main branch are set to [1, 1, 2, 3] and [4, 8, 4, 4], respectively, while those of the auxiliary branch are [4, 2, 2, 2]. For ResNet34, the main branch employs timesteps of [1, 2, 3, 3] with parallelism [8, 8, 8, 4], and the auxiliary branch adopts [2, 2, 2, 2]. As shown in \fgref{fg_TSCvsNoTSC} (a) and \fgref{fg_TSCvsNoTSC} (e), introducing intra-block parallelism significantly improves hardware utilization for both branches. Moreover, since more computational resources are allocated to latency-critical blocks, the overall latency distribution becomes more balanced across blocks, as illustrated in \fgref{fg_TSCvsNoTSC} (b) and \fgref{fg_TSCvsNoTSC} (f). Consequently, the pipeline performance improves by 7.96$\times$ for SCNN9 and 7.99$\times$ for ResNet34.

\subsubsection{Joint Evaluation of BMS+TPOS With and Without TSC}
\begin{figure*}[tbp]
    \centering
    \includegraphics[width=0.95\linewidth]{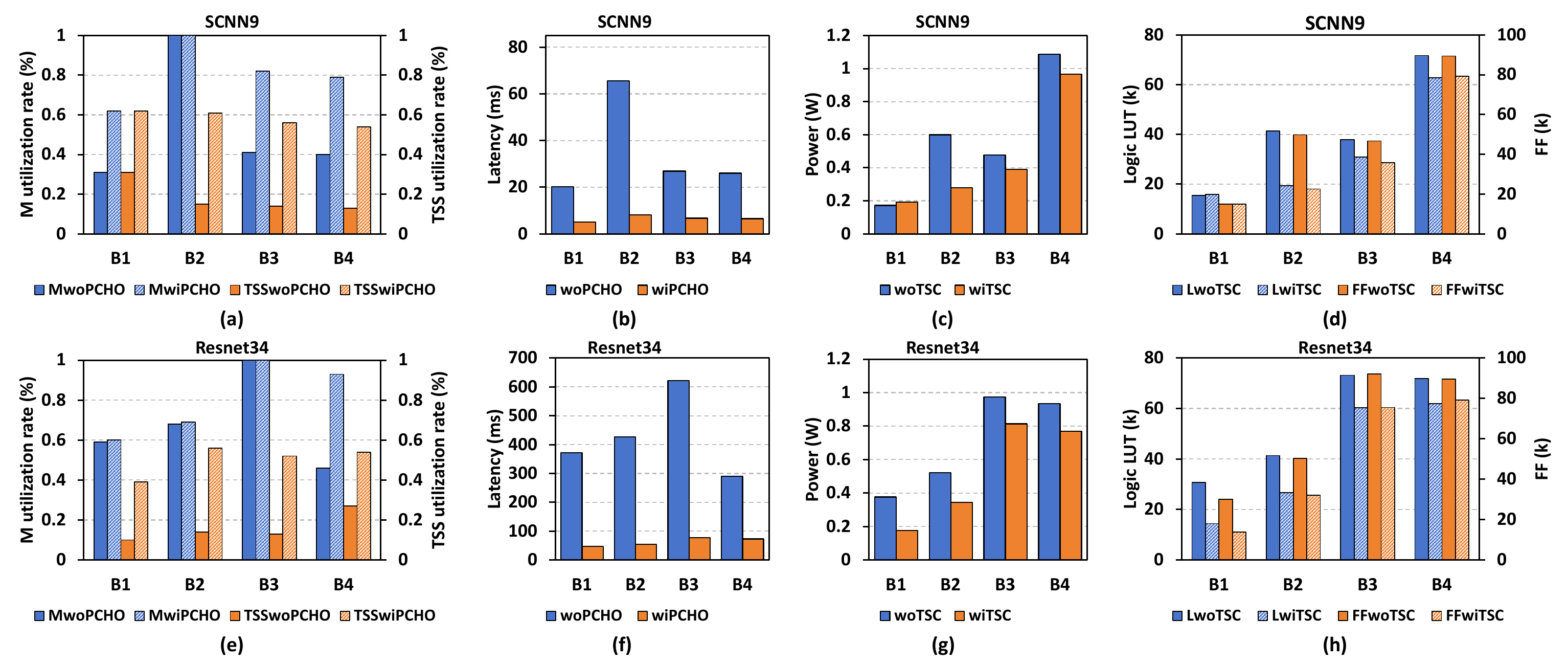}
    \caption{
    With the TPOS dataflow and TSC enabled, panels (a,e) and (b,f) report hardware resource utilization and inference latency, respectively, comparing designs with and without intra-block parallelism. With the TPOS dataflow and intra-block parallelism enabled, panels (c,g) and (d,h) report hardware power and resource usage, respectively, comparing TSC-enabled and TSC-disabled configurations.
    }
    \label{fg_TSCvsNoTSC}
    \vspace{-10pt}
\end{figure*}

\begin{figure}[tbp]
    \centering
    \vspace{-10pt}
    \includegraphics[width=0.95\columnwidth]{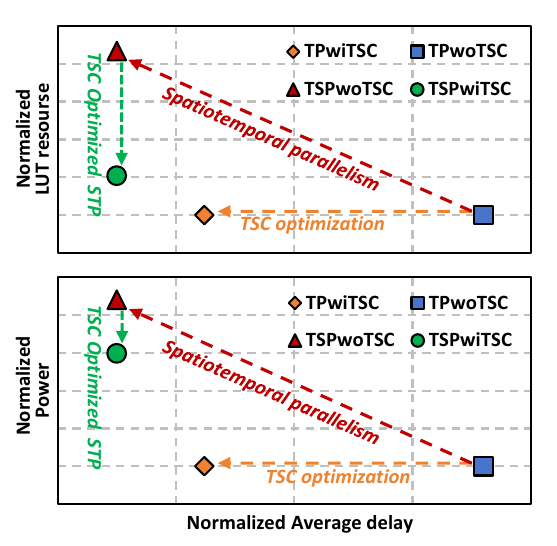}
    \vspace{-10pt}
    \caption{
    System-level normalized scatter plots (CIFAR-100 on ResNet18): (a) total LUTs versus average latency; (b) total power versus average latency.
    }
    \label{fg_tscrd}
    \vspace{-20pt}
\end{figure}

We further analyze the configurations with and without TSC when both TPOS and BMS are applied. As shown in \fgref{fg_TSCvsNoTSC}, for ResNet34 the power consumption of different blocks is reduced from 0.38-0.93W to 0.18-0.77W, yielding an average reduction of 30.4\% and a 1.5$\times$ improvement in power efficiency (\fgref{fg_TSCvsNoTSC} (g)), while for SCNN9 the average power decreases by 17.7\% with a 1.26$\times$ efficiency gain (\fgref{fg_TSCvsNoTSC} (c)). In terms of logic resources, the LUT usage of ResNet34 blocks drops from 30.77-71.93k to 14.40-61.97k, giving 13.8\%-53.2\% block-wise reductions (30.1\% on average) and about 1.5$\times$ higher logic-utilization efficiency (\fgref{fg_TSCvsNoTSC} (h)); the flip-flop (FF) usage decreases from 29.99-89.52k to 13.72-79.22k, corresponding to 11.5\%-54.2\% reductions (30.0\% on average) and about 1.53$\times$ higher FF efficiency. For SCNN9, LUT and FF utilizations are reduced on average by 20.9\% and 22.2\%, corresponding to 1.37$\times$ and 1.41$\times$ efficiency improvements, respectively.
The system-level view offers clearer insight. \fgref{fg_tscrd} presents CIFAR-100 on ResNet18 using two normalized scatter plots: total LUTs and total power versus average latency. Under spatial-temporal parallelism, TSC yields 51.92\% fewer LUTs and 22.83\% lower power, and at equal latency reduces system energy.
Overall, combining TSC with the TPOS+BMS co-optimization framework yields substantial improvements in latency balance, power efficiency, and logic resource utilization, thereby validating the proposed STISA architecture’s scalability and hardware effectiveness.

\begin{table*}[tbp]
\small                      
\vspace{-10pt}
\setlength{\tabcolsep}{3pt}
\renewcommand{\arraystretch}{1.25}
\centering
\caption{Comparison of STISA with Other SOTA Accelerators on \\ Accuracy, Throughput, Power, and Efficiency Metrics.}
\label{tab_hd1}
\begin{threeparttable}
\begin{tabularx}{\linewidth}{l|c|C|C|c|c|c|c|C|C|c}
\Xhline{1pt}
\makecell[c]{Ref} &
\makecell[c]{Device} &
\makecell[c]{Dataset} &
\makecell[c]{SNNs'\\model} &
\makecell[c]{Acc\\(\%)} &
\makecell[c]{T} &
\makecell[c]{FPS} &
\makecell[c]{kFPS$\times$MOPs\\(GOPS)} &
\makecell[c]{Power\\(W)} &
\makecell[c]{Eff\\(GOPS/W)} &
\makecell[c]{Eff/PE\\(GOPS/W/PW)} \\
\Xhline{0.5pt}
Firefly2\cite{Firefly2}          & xczu5ev   & CIFAR-10   & CIFAR-Net\tnote{1}  & 93.6    &  4     & 333.67 & 3443   & 4.9  & 702.74 & 0.09 \\
Firefly2\cite{Firefly2}          & xczu5ev   & ImageNet   & ResNet34  & 67.3    &  8     & 32.67  & 1918   & 4.9  & 391.46 & 0.05 \\
\Xhline{0.5pt}          
Firefly\cite{Firefly}            & xczu3eg   & MNIST      & SCNN5     & 98.12   &  4     & 2036   & 265.76 & 2.55 & 104.22 & 0.05 \\
Firefly\cite{Firefly}            & xczu3eg   & CIFAR-10   & SCNN7\tnote{2}     & 91.36   &  4     & 966    & 274.49 & 2.55 & 107.64 & 0.05 \\
\Xhline{0.5pt}          
DeepFire2\cite{Deepfire2}        & xcvu9p    & CIFAR-10   & VGG10-S   & 87.10   &  1     & 23255  & 10400  & 20.1 & 517.93 & /   \\
DeepFire2\cite{Deepfire2}        & xcvu9p    & ImageNet   & VGG10-L   & 40.1    &  1     & 1560   & 21100  & 47.2 & 447    & /   \\
\Xhline{0.5pt}          
Cerebron\cite{Cerebron}          & xc7z100   & MNIST      & ConvNet\tnote{3}   & 99.40   &  4     & 38461  & 40.1   & 1.4  & 28.6   & 0.11 \\
Cerebron\cite{Cerebron}          & xc7z100   & CIFAR-10   & MobileNet & 91.90   &  4     & 94.07  & 44.2   & 1.4  & 31.6   & 0.12 \\
\Xhline{0.5pt}          
Zhang et al.\cite{zhang2025power}      & xczu9eg   & CIFAR-10   & Model1\tnote{4}    & 90.5    &  4     & 827.13 & /      & 1.69 & /      & /   \\
Zhang et al.\cite{zhang2025power}      & xczu9eg   & ImageNet   & Model3\tnote{5}    & 62.3    &  4     & 44.95  & /      & 1.69 & /      & /   \\
\Xhline{0.5pt}         
E3N3\cite{e3n3}                 & xcvu13p   & CIFAR-10   & AlexNet   & 80.6    &  6     & 14.29  & /      & 4.7    & /      & /   \\
\Xhline{0.5pt}       
Li et al.\cite{li2021tcasi}     & xc7vx485t & MNIST      & MLP\tnote{6}       & 92.93   &  /     & 317    & /      & 1.6  & /      & /   \\
\Xhline{0.5pt}  
Ye et al.\cite{ye}              & xc7k325t  & SVHN       & SCNN6\tnote{7}      & 82.15   &  /    & 826.4  & 5.26  & 0.98  & 5.35      & 0.02   \\

\Xhline{0.5pt}         
\textbf{STISA}                  & xczu9eg   & CIFAR-10   & SCNN9     & 92.91   & 1.75   & 121.65 & 164.96 & 1.93 & 85.47  & 0.17 \\
\textbf{STISA}                  & xczu9eg   & CIFAR-10   & ResNet18  & 95.29   & 2      & 49.33  & 126.28 & 1.69 & 74.72  & 0.19 \\
\textbf{STISA}                  & xczu9eg   & CIFAR-100  & ResNet18  & 78.15   & 2.25   & 30.5   & 85.03  & 1.04 & 85     & 0.34 \\
\textbf{STISA}                  & xczu28dr  & ImageNet   & ResNet34  & 68.51   & 2.25   & 12.85  & 241.48 & 2.21 & 109.27 & 0.16 \\
\Xhline{1pt}
\end{tabularx}

\label{tab:gops3}
\begin{tablenotes}
\item[1]
CIFAR-Net: 32×32-32c3-256c3-256c3-mp2-256c3-256c3-256c3-mp2-512c3-mp2-1024c3-ap-10.
\item[2]
SCNN7: 32×32-16c3-64c3-p2-128c3-128c3-256c3-256c3-p2-512c3-10.
\item[3]
ConvNet: 28×28-16c3-32c3-16c3-10.
\item[4]
Model1: 32×32-16c3-64c3-p2-128c3-p2-256c3-p2-512c3-fc10.
\item[5] 
Model3: 224×224-64c7-p3-64c3-64c3-128c3-128c3-256c3-256c3-512c3-512c3-fc1000.
\item[6] 
MLP: Classification network with 728-200-100-10 structure.
\item[7] 
SCNN6: 32×32-32c3-p2-32c3-p2-256-10.
\end{tablenotes}
\end{threeparttable}
\vspace{-40pt}
\end{table*}

\begin{table*}[tbp]
\small
\vspace{-10pt}
\setlength{\tabcolsep}{1.5pt}
\renewcommand{\arraystretch}{1.25}
\caption{Resource Utilization and Comparison.}
\label{tab_hd2}
\begin{tabularx}{\linewidth}{l|c|c|c|c|c|c|c|c|c|c|c|c|c}
\Xhline{1pt}
& \makecell{Firefly2\\\cite{Firefly2}}      & \makecell{Firefly\\\cite{Firefly}}   & \makecell{DeepFire2\\\cite{Deepfire2}}     & \makecell{DeepFire2\\\cite{Deepfire2}}     & \makecell{Cerebron\\\cite{Cerebron}}  & \makecell{Zhang\\et al.\cite{zhang2025power}} & \makecell{Li et al.\\\cite{li2021tcasi}} & \makecell{E3N3\\\cite{e3n3}} &   \makecell{Ye et al.\\\cite{ye}} & \textbf{STISA} & \textbf{STISA} & \textbf{STISA} & \textbf{STISA}\\
\Xhline{0.5pt}
Device          & xczu5ev   & xczu3eg       & xcvu9p        & xcvu9p    & xc7z100   & xczu9eg & xc7vx485t & xcvu13p & xc7k325t      &xczu9eg    & xczu9eg   & xczu9eg & xczu28dr \\
\Xhline{0.5pt}
Precision       & Int8      & Int8          & Int8          & Int8      & /         & Int8      & FP16      & Int6      & FIX16     & Int8      & Int8      & Int8  & Int8\\
\Xhline{0.5pt}                  
NT & \makecell{IF/RMP} & \makecell{LIF/IF} & IF             & IF        & IF        & LIF       & LIF       & /         & EPC-LIF   & LIF       & LIF       & LIF   & LIF \\
\Xhline{0.5pt}                  
Fre             & 500       & 300           & 550           & 450       & 200       & 200       & 100       & 150       & 200       & 200       & 200       & 200   & 200 \\
\Xhline{0.5pt}                      
PE.S            & 8192      & 2304          & /             & /         & 256       & 2304      & 1094      & /         & 256       & 504       & 387       & 243   & 666 \\
\Xhline{0.5pt}                      
Av.\,LUT        & 117       & 70            & 1182          & 1182      & 277       & 274       & 303       & 1728      & 203       & 274       & 274       & 274   & 425 \\
Ud.\,LUT        & 26        & 15            & 125           & 371       & 86        & 23        & /         & 48        & 80.2      & 161.3     & 137.5     & 89    & 272.3 \\
Ut.\,LUT        & 20.5      & 21.4          & 10.6          & 31.4      & 31        & 8.4       & /         & 2.8       & 39.5      & 58.9      & 50.2      & 32    & 64 \\
\Xhline{0.5pt}              
Av.\,FF         & 234       & 140           & 2364          & 2364      & 554       & 548       & 607       & 3456      & 407       & 548       & 548       & 548   & 850 \\
Ud.\,FF         & /         & /             & /             & /         & 70.5      & 9         & /         & 50        & 138.7     & 160.4     & 137.8     & 88.1  & 270.2 \\
Ut.\,FF         & /         & /             & /             & /         & 12.7      & 1.6       & /         & 1.4       & 34.1      & 29.3      & 25.1      & 16.1  & 31.8  \\
\Xhline{0.5pt}              
Av.\,BRAM       & 133       & 216           & 2160          & 2160      & 755       & 912       & 1030      & 2688      & 445       & 912       & 912       & 912   & 1080 \\
Ud.\,BRAM       & 8         & 162           & 511           & 1757      & 283       & 224       & /         & /         & 245.5     & 160       & 160       & 160   & 230 \\
Ut.\,BRAM       & 6         & 75            & 23.7          & 81.3      & 37.5      & 24.6      & /         & /         & 55.2      & 17.5      & 17.5      & 17.5  & 21.3 \\
\Xhline{0.5pt}              
Av.\,DSP        & 1248      & 360           & 6840          & 6840      & 2020      & 2520      & 2800      & 12288     & 840       & 2520      & 2520      & 2520  & 4272 \\
Ud.\,DSP        & 512       & 288           & 2025          & 5400      & 0         & 0         & 0         & 0         & 0         & 0         & 0         & 0     & 0 \\
Ut.\,DSP        & 41        & 80            & 29.6          & 79        & 0         & 0         & 0         & 0         & 0         & 0         & 0         & 0     & 0 \\
\Xhline{1pt}
\end{tabularx}

\label{tab:gops3}
    \begin{tablenotes}
         \item[] NT: Neuron Type, Fre:  Frequency (MHz), PE.S: PE Array Size, Av.LUT: Available LUT(k), Ud.LUT: Used LUT (k), Ut.LUT: Logic LUT Utilizaiton (\%), Av.FF: Available FF(k), Ud.FF: Used FF (k), Ut.FF: FF Utilizaiton (\%), Av.BRAM: Available BRAM, Ud.BRAM: Used BRAM, Ut.BRAM: BRAM Utilizaiton (\%), Av.DSP: Available DSP, Ud.DSP: Used DSP.
\end{tablenotes}
\vspace{-10pt}
\end{table*}

\subsection{Comparison and Analysis with Previous Work}
\tabref{tab_hd1} and \tabref{tab_hd2} compare the performance of STISA with other SOTA FPGA-based SNN accelerators using identical benchmark datasets (CIFAR-10, CIFAR-100, and ImageNet). The comparison encompasses key metrics including classification accuracy, throughput, accelerator efficiency, and hardware resource utilization. Although the evaluated accelerators adopt different SNN mapping strategies (e.g., streaming or overlap), a consistent experimental setup ensures a fair and meaningful comparison. Following the methodology in \cite{Deepfire2,Firefly2}, the total operation count of each SNN model is derived by computing the equivalent ANN FLOPs using established profiling tools such as $ptflops$.
It should be emphasized that DeepFire2 and Firefly2 do not support any form of sparsity exploitation; therefore, their inference performance depends solely on the total number of operations, simulation timesteps, and operating frequency. As the computational workload scales linearly with the number of timesteps, it is quantified as the product of operations and timesteps (Ops × T). Based on the data summarized in \tabref{tab_hd1}, several key observations can be drawn.

\noindent\textbf{1) Latency-Accuracy Trade-off:}
DeepFire2 prioritizes inference latency by adopting an ultra-short simulation setting (T = 1), which inevitably compromises classification accuracy. In contrast, Firefly2 achieves higher accuracy on more complex datasets such as ImageNet. By integrating the proposed TSC mechanism, STISA significantly reduces the average timesteps of the main branch while maintaining high accuracy, achieving 95.29\% on CIFAR-10 and 68.51\% on ImageNet. These results are comparable to the SOTA performance reported in recent SNN literature.

\noindent\textbf{2) Performance and Portability Considerations:}
Although STISA exhibits slightly lower FPS and energy efficiency than Firefly2 and DeepFire2 on ImageNet, these designs rely on DSP-centric optimization strategies that are tightly coupled to specific FPGA architectures. As acknowledged in the Firefly publication, such designs elevate the maximum achievable clock frequency but limit Verilog portability and complicate migration to ASIC implementations. While DSP utilization can enhance SNN acceleration performance and energy efficiency, careful tuning of parallelism and bit-width is required to avoid being constrained by DSP availability or underutilizing high-precision resources. In contrast, STISA is implemented without dependence on DSP-specific features, and still achieves higher per-PE energy efficiency.

\noindent\textbf{3) Hardware Scalability and Deployability:}
DeepFire2 depends on large, high-end FPGA platforms, which are impractical for embedded or edge computing scenarios. To ensure a fair comparison of energy efficiency and resource utilization under equivalent workloads, STISA is configured with 243 PEs for the ResNet18 network on the CIFAR-100 dataset. This configuration allows for direct evaluation of its scalability across different network sizes.

\noindent\textbf{4) Architectural Flexibility:}
STISA demonstrates strong architectural flexibility, supporting variable model sizes and diverse network topologies. The accelerator enables independent parallelization factors for different convolutional layers, effectively improving PE utilization and hardware efficiency. Such adaptability highlights STISA’s suitability for heterogeneous SNN workloads and various deployment scenarios.

\section{Conclusion} \label{conclusion}
In this work, STISA, a unified algorithm-hardware co-design framework, is proposed for low-latency and energy-efficient SNN inference under stringent resource constraints. On the algorithm side, the proposed TSC performs block-wise adaptive timestep allocation, reducing synaptic operations by 24.44\%-51.71\% while maintaining competitive accuracies of 96.38\%, 81.10\%, 68.51\%, and 82.40\% on CIFAR-10, CIFAR-100, ImageNet, and DVS-CIFAR10, respectively. On the hardware side, STISA develops a spatio-temporal parallel architecture that combines a TPOS dataflow and a BMS architecture to eliminate membrane-potential buffering, improve weight and state locality, and enable both inter- and intra-block parallelization with bandwidth-aware resource allocation. For ImageNet on ResNet34, the FPGA prototype achieves up to 7.99$\times$ speedup, 30.4\% power and 30.1\% LUT reductions, and 0.16~GOPS/W/PE, demonstrating an efficient latency-energy-accuracy trade-off for SNN inference under limited resources.



\vspace{-10mm} 

\begin{IEEEbiography}[{\includegraphics[width=1in,height=1.25in,clip,keepaspectratio]{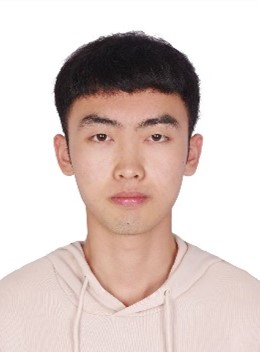}}]{Kainan Wang}
received the B.S. degree in Electronic and Information Engineering from the School of Information Science, Yunnan University, Kunming, China, in 2022. He is currently pursuing the Ph.D. degree in electronics science and technology with the College of Information Science and Electronic Engineering, Zhejiang University, Hangzhou, China. His current research interests include hardware-algorithm co-design for neuromorphic computing and domain-specific processor architectures for edge applications.
\end{IEEEbiography}

\vspace{-10mm} 

\begin{IEEEbiography}[{\includegraphics[width=1in,height=1.25in,clip,keepaspectratio]{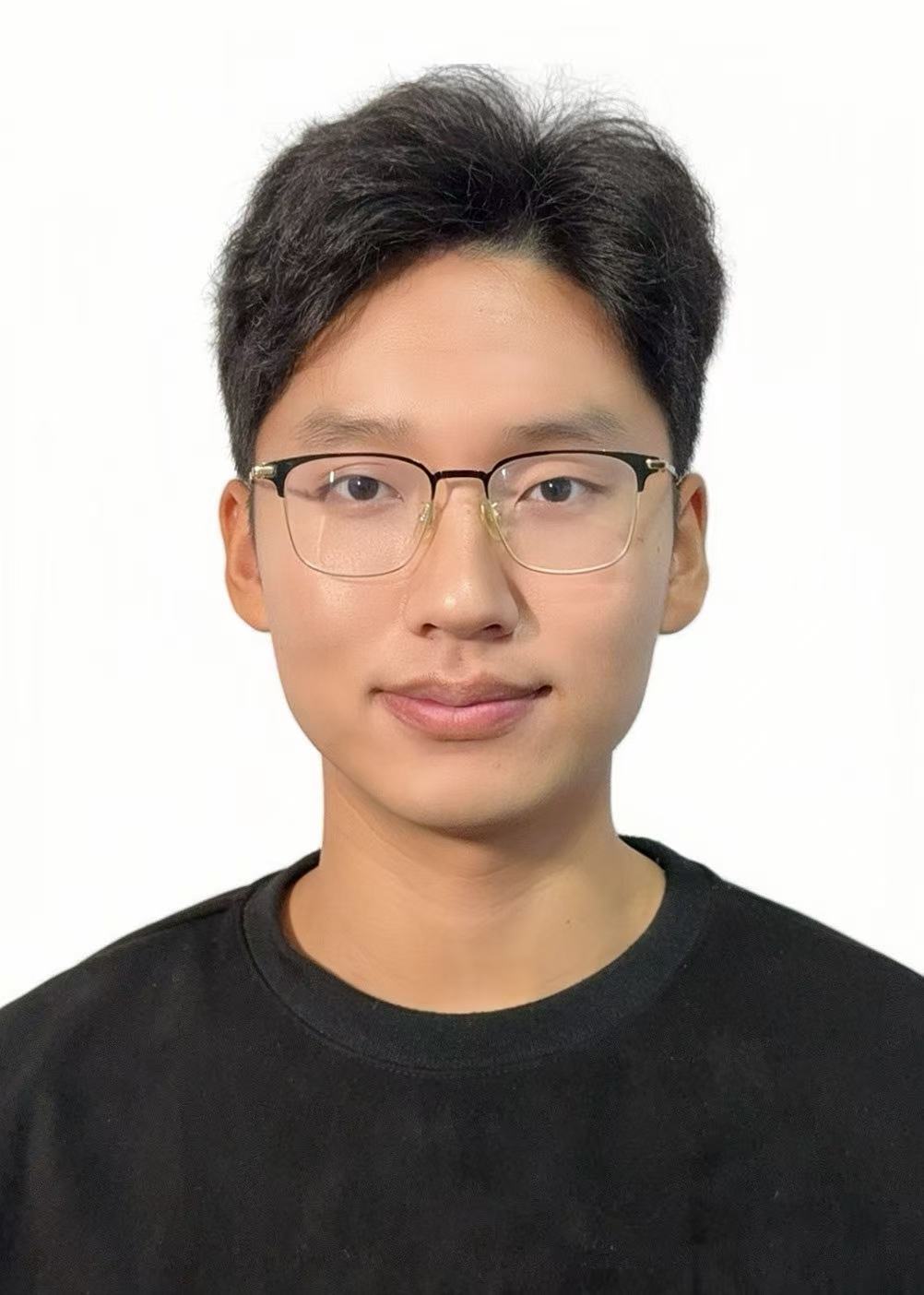}}]{Chengyi Yang}
received the M.S. degree in Electronic and Information Engineering from Zhejiang University, Hangzhou, China, in 2025.
He is currently an IC Design Engineer with HiSilicon Technologies Co., Ltd., Hangzhou, China. His research interests include computer architecture, AI accelerators, and heterogeneous computing systems (including GPUs and NPUs).
\end{IEEEbiography}

\vspace{-10mm}  

\begin{IEEEbiography}[{\includegraphics[width=1in,height=1.25in,clip,keepaspectratio]{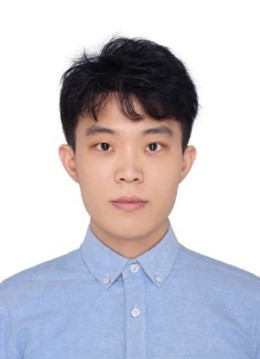}}]{Lei Liu}
received the B.S. degree in marine technology from Wuhan University of Technology, Wuhan, China, in 2019. He is currently pursuing the M.S. degree in artificial intelligence with Zhejiang University-University of Illinois Urbana-Champaign Institute, Zhejiang University, Haining, China. His current research interests include neuromorphic computing and high-performance training algorithms for spiking neural networks.
\end{IEEEbiography}

\vspace{-10mm} 

\begin{IEEEbiography}[{\includegraphics[width=1in,height=1.25in,clip,keepaspectratio]{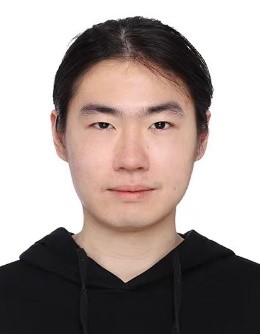}}]{Chengting Yu}
received the B.S. degree from Zhejiang University, Hangzhou, China, in 2021, where he is currently pursuing the Ph.D. degree with the College of Information Science and Electronic Engineering. His research interests include deep learning and neuromorphic computing, centered on neural network architecture design and efficient (lightweight) training algorithms.
\end{IEEEbiography}

\begin{IEEEbiography}[{\includegraphics[width=1in,height=1.25in,clip,keepaspectratio]{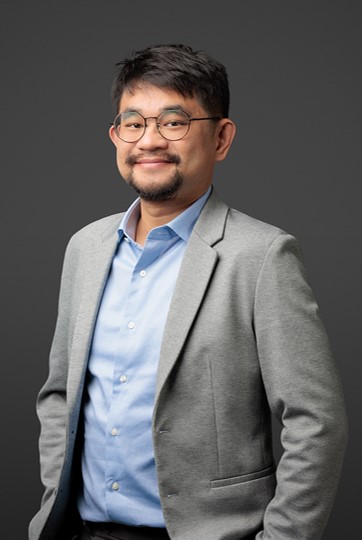}}]{Yee Sin Ang}
received the B. Med. and Radiation Physics (Hons.) degree in 2010 and the Ph.D. degree in Theoretical Condensed Matter Physics in 2014, both from the University of Wollongong, Wollongong, Australia. He is currently an Assistant Professor with the Science, Mathematics and Technology (SMT) Cluster, Singapore University of Technology and Design (SUTD), Singapore. He was an alumnus of the 69th Lindau Nobel Laureate Meeting, Lindau, Germany, in 2019. His research focuses on the physics and design of nanoscale devices for energy-efficient computing and renewable energy technologies. His research interests include first-principles modeling of two-dimensional (2D) materials and heterostructures, metal contacts to 2D semiconductors, charge injection and transport at low-dimensional and topological interfaces, ultrascaled device design, beyond-CMOS electronics (e.g., valleytronics and spintronics), and light–matter interactions and GHz/THz photonics.
\end{IEEEbiography} 

\begin{IEEEbiography}[{\includegraphics[width=1in,height=1.25in,clip,keepaspectratio]{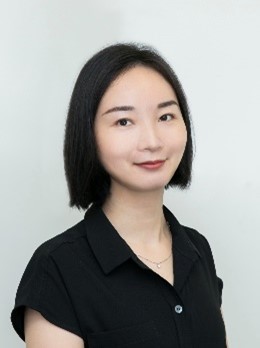}}]{Bo Wang}
(Senior Member, IEEE) received the Ph.D. degree in Electrical and Electronic Engineering from Nanyang Technological University, Singapore, in 2015. From 2015 to 2016, she worked as a Staff Circuit Design Engineer at MediaTek, Singapore. She then joined the National University of Singapore as a Post-Doctoral Research Fellow from 2016 to 2020. Since 2020, she has been an Assistant Professor at the Singapore University of Technology and Design, Singapore. She has authored and co-authored many papers published in prestigious journals and conference proceedings, including TNNLS, JSSC, TCAS-I, TVLSI, TCAS-II, A-SSCC, DAC, DATE, ISLPED, and MobiSys. Her research interests span various aspects of energy-efficient system, architecture and circuit design, as well as design automation tools for hardware–software co-optimization.

Dr. Wang received the Distinguished Design Award at the IEEE Asian Solid-State Circuits Conference (A-SSCC) in 2023. She was also a recipient of the Best Paper Award at the ACM/IEEE International Symposium on Low Power Electronics and Design (ISLPED) in 2023 and the International SoC Design Conference (ISOCC) in 2024 and 2014. She currently serves as an Associate Editor for the IEEE Open Journal of Circuits and Systems and has served as a Guest Editor for Frontiers in Neuroscience.
\end{IEEEbiography} 

\begin{IEEEbiography}[{\includegraphics[width=1in,height=1.25in,clip,keepaspectratio]{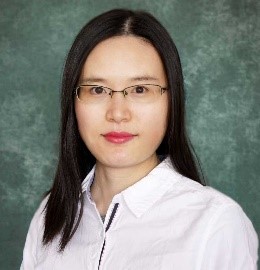}}]{Aili Wang}
(Member, IEEE) received the Ph.D. degree in electrical and computer engineering from the University of Washington, Seattle, WA, USA, in 2019. Since 2020, she has been an Assistant Professor with the Zhejiang University-University of Illinois Urbanna-Champaign Institute, Zhejiang University, Haining, China. Her research interests include low power analog and mixed signal integrated circuits design and the codesign of algorithm and hardware for brain inspired computing. She has authored and co-authored many papers published in prestigious journals and conference proceedings, including TCAS-I, TVLSI, Sensors Journal, TAI, NCE, CICC, ISCAS, NeurIPS, ICML, CVPR, and ECAI. Dr. Wang is a TPC member of CSTIC since 2022 and a TPC member of ICICM since 2023, She was the recipient of IEEE Solid-Sate Circuits Society Student Travel Award in 2018.
\end{IEEEbiography}

\end{document}